\documentclass[11pt,a4paper]{article}
\usepackage{jheppub}

\usepackage[T1]{fontenc}
\usepackage{placeins}

\usepackage{amsmath}
\usepackage{amssymb}
\usepackage{mathtools}
\usepackage{amsthm}
\usepackage{physics}
\usepackage{bm}
\usepackage{slashed,accents}

\usepackage{graphicx}
\usepackage{tikz-cd}
\usepackage{tikz-feynman}
\tikzfeynmanset{compat=1.1.0}
\definecolor{allOrderBlue}{rgb}{0.4,0.5,1}

\usepackage{enumitem}
\usepackage[dvipsnames]{xcolor} % for draft note colours
\usepackage[nameinlink,noabbrev]{cleveref}

\newcommand{\dhat}{\hat{d}}

\newcommand{\integrand}{H}
\newcommand{\waveA}{\mathcal{A}}
\newcommand{\waveB}{\mathcal{B}}

\newcommand{\obsCoord}{x}

\newcommand{\Arad}{A_\textrm{rad}}
\newcommand{\Jschild}{J_\textrm{cons}}

\newcommand{\polvector}{\varepsilon}

\newcommand{\rSchild}{\mathring{r}}
\newcommand{\uSchild}{\mathring{u}}
\newcommand{\aSchild}{\mathring{a}}
\newcommand{\vrSchild}{\mathring{\vec{r}}}

\newcommand{\vhrSchild}{\mathring{\hat{\vec{r}}}}
\newcommand{\vhvSchild}{\mathring{\hat{\vec{v}}}}

\renewcommand{\[}{\begin{equation}\begin{aligned}}
\renewcommand{\]}{\end{aligned}\end{equation}}

\renewcommand{\vec}[1]{\boldsymbol{#1}}

\newcommand{\zpc}[3]{z_{#1}^{{#2},({#3})}}
\newcommand{\dzpc}[3]{\dot{z}_{#1}^{{#2},({#3})}}
\newcommand{\ddzpc}[3]{\ddot{z}_{#1}^{{#2},({#3})}}

\newcommand{\Atilde}{\tilde{A}}

\DeclareMathOperator\sgn{sgn}
\newcommand{\delh}{\hat{\delta}}

\newcommand{\tddot}[1]{{{#1}^{\prime\prime}}}
\newcommand{\tdddot}[1]{{{#1}^{\prime\prime\prime}}}
\tikzset{
  onshell/.style={
    decoration={
      markings,
      mark=at position #1 with {
        \draw[white, line width=5pt] (0,-0.25) -- (0,0.25);
        \draw[dashed] (0,-0.25) -- (0,0.25);      }
    },
    postaction={decorate}
  }
}

\begin{document}

\title{
From A to B: Amplitudes for Bound Observables 
}
\author[a]{Majed Khalaf,}
\emailAdd{majed.khalaf@mail.huji.ac.il}
\affiliation[a]{Racah Institute of Physics, Hebrew University of Jerusalem, Jerusalem 91904, Israel}

\author[b]{David A. Kosower,}
\emailAdd{David.Kosower@ipht.fr}
\affiliation[b]{Institut de Physique Th\'{e}orique, CEA, CNRS,
Universit\'{e} Paris-Saclay, F-91191 Gif-sur-Yvette cedex, France}

\author[c]{Donal O'Connell,}
\emailAdd{donal@ed.ac.uk}
\affiliation[c]{Higgs Centre for Theoretical Physics, School of Physics and Astronomy,
The University of Edinburgh, Edinburgh EH9 3JZ, Scotland, UK}

\author[a]{Ofri Telem,}
\emailAdd{t10ofrit@gmail.com}

\author[c]{and Shen Yan}
\emailAdd{S.Yan-14@sms.ed.ac.uk}

\abstract{
We set up a relativistically covariant perturbative formalism for the computation of classical bound-state observables.
Our method applies when the bound bodies are sufficiently far apart.
Focusing on a proof of concept in electromagnetism, we demonstrate that waveforms are integrals over a rational function which is closely related to the integrand of standard scattering amplitudes.
We determine the next-to-leading order electromagnetic waveform in an inspiral using our methods.
}

\maketitle
\newpage

\def\draftnote#1{\textit{#1}}

\section{Introduction}

The first decade of observing gravitational-wave emission
from coalescing binaries by the LIGO--VIRGO--KAGRA  collaboration (LVK) has given us a rich data set.
It is composed mostly of binary black hole mergers, with a few black hole--neutron star and binary neutron star mergers as well.  
The second decade now underway promises even more, with continuing observations by LVK; third-generation terrestrial observatories coming online towards its end; and perhaps even the budding of space-based missions.

The first decade has also witnessed a flowering of theoretical developments, in applications of ideas and methods from quantum scattering amplitudes to classical gravitational-wave physics. 
These include several different approaches. 
One applies calculations of the conservative potential~\cite{Bern:2019nnu,Mogull:2020sak,Kalin:2020fhe,Bern:2020buy,Jakobsen:2021smu,Bern:2021yeh,Bern:2021dqo,Jakobsen:2022psy,Jakobsen:2023ndj,Jakobsen:2023hig,Jakobsen:2023pvx,Bern:2024adl,Driesse:2024feo,Driesse:2024xad,Bern:2025zno,Bern:2025wyd,Driesse:2026qiz,Dlapa:2026oyq} and matching~\cite{Neill:2013wsa,Cheung:2018wkq} to an effective field theory~\cite{Goldberger:2004jt} in order to fix parameters in a relativistic Hamiltonian such as that in an Effective One Body~\cite{Buonanno:1998gg} approach.
Another pursues direct calculations of waveforms~\cite{Kosower:2018adc,Cristofoli:2021vyo,Adamo:2022qci,DeAngelis:2023lvf,Aoude:2023dui,Brandhuber:2023hhl,Herderschee:2023fxh,Brandhuber:2023hhy,Georgoudis:2023lgf,Elkhidir:2023dco,Bohnenblust:2023qmy,Adamo:2024oxy,Bini:2024rsy,Brunello:2024ibk,Bohnenblust:2025gir,Brunello:2025eso,Aoki:2026eos,Aoki:2026fgt}.
Other approaches are based on the assumption of an eikonal form~\cite{DiVecchia:2020ymx,DiVecchia:2021bdo}.
For some recent reviews, see Refs.~\cite{Travaglini:2022uwo,Kosower:2022yvp,Bern:2022wqg,Bjerrum-Bohr:2022blt,DiVecchia:2023frv}.

Amplitude-based waveform calculations have mostly worked
in the scattering regime, with applications to bound
states via indirect methods. 
One approach is to match scattering amplitudes to a two-particle effective theory which treats interactions via a non-local potential~\cite{Neill:2013wsa,Cheung:2018wkq}.
This effective theory can then be used to compute bound-state observables --- though there are caveats.
An alternative, known as the boundary-to-bound or B2B map, involves a specific analytic continuation of certain observables directly from scattering to bound motion~\cite{Kalin:2019rwq,Kalin:2019inp,Cho:2021arx,Herrmann:2021lqe}.

At present, methods of transferring information from the scattering to the bound case have limits on their applicability. 
For example, the B2B map successfully relates some observables in the  scattering and bound domains, with restrictions to large eccentricity orbits for so-called tail contributions~\cite{Cho:2021arx}.
It would be desirable to relax this limitation, and explore a more general approach.
The procedure of matching to a potential is intrinsically non-radiative, but extending the basic idea of a non-local effective theory to radiative cases is possible.
This leaves a problem: the (time-local) interaction Hamiltonian relevant for scattering encounters differs at higher orders from that relevant for bound motion~\cite{Bini:2017wfr,Bini:2024tft}.
These differences arise because gravitational radiation emitted at some time continues to interact with the source at later times: interactions which differ in unbound and bound contexts.
As a result, a direct on-shell computation of bound-state quantities at comparable masses, and especially the waveforms of direct interest to observers, has remained elusive. 
A different expansion that arises in the Extreme-Mass-Ratio regime does allow a wealth of direct analytic computations of orbital frequencies and spin precession~\cite{Gonzo:2023goe,Bini:2018ylh}, and high-order post-Newtonian self-force corrections for circular and eccentric Schwarzschild orbits~\cite{Bini:2014nfa,Hopper:2015icj}, radiative fluxes~\cite{Munna:2020iju,Khalaf:2023ozy,Castillo:2024isq,Warburton:2024xnr,Khalaf:2025jpt,Capozziello:2025tfi,Khalaf:2026ovs}, and inspiral waveforms and their memory~\cite{Isoyama:2021jjd,Cunningham:2024dog}.
This suggests that a similar connection between unbound and bound regimes should exist in the ordinary perturbative expansion.

A promising alternative avenue of investigation in the comparable-mass regime was suggested earlier this year by Gonzo and Mogull~\cite{Gonzo:2026yha}.
Their suggestion was to combine the worldline formalism and the Magnus expansion of the finite-time evolution operator $U(t, t_0)$ in a quantum-mechanical formalism.  
Retaining explicit time dependence allows the same expansion to describe both scattering and bound motion.

In this paper we propose a different approach to relating scattering and bound observables.
We construct a covariant perturbative framework for bound motion. 
We further formulate an algorithm for extracting bound-state waveform integrands from scattering amplitudes.
As a proof of principle, we apply the method to electrodynamics and compute the waveform of a quasi-circular inspiral through next-to-leading order (NLO) in our perturbative expansion.

The gravitational systems which motivate our construction are well-separated binaries, with orbital radii large compared to their Schwarzschild radii.
This is the weak field regime, so a perturbative approach is appropriate.
Binaries of this type emit radiation slowly compared to their orbital period.
Accordingly, the orbital radii decrease adiabatically.

While we focus on bound systems, we also wish to preserve one of the successes of the on-shell methods of amplitudes applied to gravitational wave physics: manifest Lorentz invariance.
Scattering amplitudes have been developed for application to high-energy colliders, so it is very natural to write them in terms of Lorentz scalars.
Applied to interaction potentials, for example, the result at fixed order in Newton's constant $G$ retains the full velocity dependence.
This relativistic perturbative organisation is sometimes known as the post-Minkowskian expansion in classical gravity.
This is in contrast to the post-Newtonian expansion, which involves fixed orders of both Newton's constant \emph{and} velocity.
The expansion of a relativistic perturbative expression then yields an infinite series of post-Newtonian terms.
This series of post-Newtonian powers is often the most useful output of amplitudes computations applied to gravitational-wave physics.

There is some tension between our goals of using a relativistic formalism and describing bound-state physics in a weak field.
The virial theorem $2GM/r \sim v^2$ (in units where the speed of light is unity) links the coupling and velocity expansions: 
weak fields imply small velocities.
While we could use a non-relativistic formulation, a relativistic approach allows us to access all the velocity corrections at a given power of $G$.
For this reason, we use a relativistic formulation throughout.
Care is needed when interpreting our results because terms at higher order in the coupling, but lower order in the velocity, may be numerically larger than terms we retain.

A first challenge is to find a reference bound trajectory which can be the basis of a perturbative expansion.
Whatever this zeroth-order bound orbit is, it must be simple enough to admit an exact treatment and still be somehow ``close'' to the physical bound motions of interest.
However, emission of radiation is always a feature of relativistic dynamics, leading to the gradual decay of bound classical orbits and to rich dynamics.
A way out is to suppose that the loss of energy due to emission is precisely balanced by incoming radiation.
This balancing happens to be the starting point for the Wheeler--Feynman absorber theory~\cite{Wheeler:1945ps,Wheeler:1949hn}.
We adopt the Wheeler--Feynman boundary conditions for our zeroth-order bound motion, including physical radiative effects order-by-order in perturbation theory.
Put simply, our zeroth order trajectories will be exact solutions of the equations of motion, but with unphysical boundary conditions.
The physical motion, satisfying the usual retarded boundary conditions, is obtained by correcting the Wheeler--Feynman solution to include radiative effects order by order.
We will describe our perturbative setup in more detail in Section~\ref{sec:pertSetup}.

Although our approach is quite general, we make two simplifications at the outset. 
Firstly, we focus on electrodynamics as a proof of principle.
Secondly, we consider a precisely circular solution to the time-symmetric Wheeler-Feynman equations of motion due to Schild~\cite{Schild:1963}.
We expect that it will be possible to extend our approach to gravity~\cite{Hoyle:1964,Friedman:2005rx}, leaving the details to future work.

The rest of our paper is as follows. 
As mentioned above, we describe how our perturbation theory is organised in more detail in Section~\ref{sec:pertSetup}.
The topic of Section~\ref{sec:powerCounting} is the nature of the parameter suppressing higher order corrections in our approach.
Section~\ref{sec:frules} assembles the propagators and vertices needed to compute waveforms.
In Section~\ref{sec:unbound} we explain in detail how the integrands of our bound-state perturbative approach are related to ordinary flat-space amplitudes.
We further describe an algorithm for extracting bound integrands from standard amplitudes.
Section~\ref{sec:results} describes the computation of the electromagnetic waveform at next-to-leading order in our perturbation theory.
We close the main text of the article with a discussion in Section~\ref{sec:concl}.
The Appendices provide an overview of the Schild solution, and an example of some manipulations required to build bound-state integrands from ordinary scattering amplitudes.

\section{Perturbation Theory for Bound Orbits}
\label{sec:pertSetup}

We begin by setting up our perturbative computation. 
Our interest is in the radiation field generated by compact bound systems (binaries of bodies such as black holes or neutron stars) focusing on times early in their inspiral so that the effect of radiation on the trajectories can be treated as a perturbation to a dominant stable orbit.
As discussed above, we treat bound electromagnetic systems in this article, postponing a discussion of the gravitational case to future work.

Our starting point is a point-particle action,
\[\label{eq:startingAction}
I[A, x_i] = - \frac14 \int \dd^4 x \, F_{\mu\nu} F^{\mu\nu} - \sum_i \int \dd \tau_i \Big( \tfrac12 m_i \dot x_i^2 + eQ_i \, A(x_i) \cdot \dot x_i  \Big) \,,
\]
augmented by the standard action for the electromagnetic field. The sum runs over the two point particles $i = 1, 2$; the overdots indicate proper time $\tau_i$ derivatives; $F^{\mu\nu}$ is the electromagnetic field strength, the electric coupling $e$ is the charge of the proton, and the particle charges\footnote{We work in Lorentz-Heaviside units, setting also $c=1$.} are $eQ_i$. 
In an application to scattering processes we would expand the trajectories $x_i(\tau_i)$ about an unbound straight-line trajectory
\[\label{eq:scatteringTrajectory}
x_i^\mu = b_i^\mu + u_i^\mu \tau_i + z^\mu_i(\tau_i) \,,
\]
which is parameterized by unperturbed proper velocities $u_i$ and spatial offsets $b_i$. 
The functions $z_i(\tau_i)$ are the fluctuations about the unperturbed trajectories.
One obtains the action for $z_i$ by substituting this expression into the action~\eqref{eq:startingAction}, obtaining interaction vertices as a power series in derivatives of $A_\mu$.
The observables obtained from this action are the same as those obtained from the classical limit of standard amplitudes techniques, such as BCFW~\cite{Britto:2005fq}, KMOC~\cite{Kosower:2018adc} and so on.

We seek to mimic the separation between a dominant trajectory $r_i(\tau_i)$, and perturbations in a more general setting, and so 
we write,
\[\label{eq:totalTrajectory}
x_i(\tau_i) = r_i(\tau_i) + z_i(\tau_i) \,.
\]
In the scattering case, we will take,
\[\label{eq:scatteringVacuum}
r_i^\mu(\tau_i) = b_i^\mu + u_i^\mu \tau_i \,.
\]
Focusing on point-particle terms which are dependent on $A_{\mu}$ and $z_i$, 
integrating  the mixed $\dot z\cdot \dot r$ term by parts, 
and adding the following total derivative,
\begin{equation}
 	\sum_i e Q_i\int d\tau_i\;\frac{d}{d\tau_i} \sum_{n=1}^\infty
	\frac1{n!} z_i^{\nu_1}\cdots 
    z_i^{\nu_n}\partial_{\nu_1}\cdots\partial_{\nu_{n-1}} A_{\nu_n}(r_i)
	\label{TotalDerivative1}
\end{equation}
the action~\eqref{eq:startingAction} becomes,
\begin{eqnarray}\label{eq:expandedAction}
I[A, z_i] &=& - \frac14 \int \dd^4 x \, F_{\mu\nu} F^{\mu\nu} -\sum_i  \int \dd \tau_i \,e Q_i \, A(r_i) \cdot \dot{r}_i \nonumber\\[5pt]
&&-\sum_i \int \dd \tau_i \biggl\{ \tfrac12 m_i \dot z_i^2 - z_{i\,\mu} \big( m^{\vphantom{\mu}}_i  \ddot r^\mu_i - eQ_i \, F^{\mu\nu}(r_i) \dot r_{i\,\nu}\big)\nonumber\\[5pt]
&&\hspace*{20mm} 
+ \frac{e Q_i}{2}z_{i\,\mu}\bigl(z_i^\alpha \partial_\alpha F^{\mu\nu}(r_i)\dot{r}_{i\nu}+F^{\mu\nu}\dot{z}_{i\nu}\bigr)
\\\nonumber&&\hspace*{20mm}+e Q_i\sum_{n=3}^\infty \frac{(n-1)}{n!}
                z_i^{\nu_1}\cdots z_i^{\nu_{n-1}} \dot z_i^{\nu_{n}}
                     \partial_{\nu_1}\cdots \partial_{\nu_{n-2}} F_{\nu_{n-1}\nu_n}(r_i)
\\\nonumber&&\hspace*{20mm}+e Q_i\sum_{n=3}^\infty 
  \frac1{n!} z_i^{\nu_1}\cdots z_i^{\nu_n}
\partial_{\nu_1}\cdots \partial_{\nu_{n-1}} F_{\nu_n\mu}(r_i)\,\dot r_i^\mu 
\biggr\} \,.
\end{eqnarray}
Here, double overdots indicate a second derivative with respect to proper time.
We have chosen to integrate by parts so that the action does not involve $\ddot r$ beyond the one appearance visible in Eq.~\eqref{eq:expandedAction}.
The worldline Feynman rules we will list below follow from this particular form of the action.

Let's now look at terms which are linear in $z_i{}_\mu$. 
Of these, the second term in parentheses is manifestly suppressed by a power of $e$ but for generic $r_i$ the first term is not.
We want to choose $r_i$ that are close to the physical trajectories of interest, have a closed form solution, and suppress the first term by a power of $e$ as well.
These requirements make it likely that such an $r_i$ must be non-radiating.

To be precise, we define the conservative Green's function to be the half-retarded plus half-advanced Green's function:
\[\label{eq:consProp}
G_\textrm{cons}(x) &= \frac{1}{2} \left(G_\textrm{ret}(x)+ G_\textrm{adv}(x) \right) \,.
\]
In the present article we will consider trajectories which are close to circular orbits.
It turns out that we can satisfy the requirements above while maintaining a manifestly relativistic form by choosing the Schild solution~\cite{Schild:1963}.
This is an exact closed-form circular solution of the Wheeler--Feynman electromagnetic two-body problem, using the time-symmetric Green's function~\eqref{eq:consProp}.
Because it is non-radiating we will refer to it as the ``conservative'' solution.
When we embed such a solution in the physically--relevant retarded electrodynamic theory, there \emph{is} radiation --- which we compute perturbatively.

The Wheeler--Feynman equations for the two--body problem build a conservative field strength from the solution of the Maxwell equation,
\[\label{eq:FWmaxwell}
A^{\mu}_\textrm{cons} (\obsCoord) = \int d^4y \, G_\textrm{cons} (\obsCoord-y) \Jschild^\mu(y) \,.
\]
The current $\Jschild^\mu$ is generated by both particles' motion,
\[\label{eq:Schildcurrent}
\Jschild^\mu(x) = \sum_i eQ_i \int d\tau_i \, \dot{r}_i^\mu(\tau_i) \, \delta^4(x- r_i(\tau_i)) \,,
\]
where $r_i$ are generic non-radiating point-particle trajectories given by solving the equations,
\[\label{eq:conservativeEOM}
m_i \frac{d^2}{d\tau_i^2}{r}^\mu_i(\tau_i) &= eQ_i \, F^{\mu\nu}_\textrm{cons}(r_i) \frac{d}{d\tau_i} r_{i\nu}  \, .
\]
In computing $F^{\mu\nu}_\textrm{cons}(r_j)$ we omit the self-energy contribution\footnote{As the particles do not radiate, there is no need to include a radiation-reaction force. In the classical approximation, this is the only aspect of the self-field that needs to be included.} $i = j$ from Eq.~\eqref{eq:Schildcurrent}.

It may be helpful to think of a different physical picture for the solution in Wheeler--Feynman electrodynamics.
We start with conventional classical electrodynamics, with its retarded boundary conditions, and add in the radiation from infinity needed to precisely cancel the radiation emitted by the orbiting binary.
The electromagnetic field is in a standing wave, and therefore does not transmit energy to infinity.
Accordingly, the binary does not lose energy, and the particles can stay in circular orbits.

At present, an explicitly-known Wheeler--Feynman solution is the circular and planar one due to Schild \cite{Schild:1963}. This solution has particularly simple trajectories of the form,
\begin{eqnarray}\label{eq:schildStarter}
\rSchild^\mu_i(\tau)=\left(\gamma_i \tau,\pm\Omega^{-1}v_i\cos\left[\gamma_i \Omega\tau\right],\pm\Omega^{-1}v_i\sin \left[\gamma_i \Omega\tau\right],0\right)\,,
\end{eqnarray}
where we choose the $+$ ($-$) sign in~\eqref{eq:schildStarter} for $\rSchild_1$ ($\rSchild_2$). 
The quantities $\gamma_i$ are the usual Lorentz boost factors $\gamma_i^2 = 1/(1 - v_i^2)$, $\Omega$ is a common angular frequency with respect to the time coordinate of the binary's rest frame, and $v_i$ are coordinate velocities of each body.
The reader may find further discussion of the Schild solution in Appendix~\ref{app:Schild}.
We denote quantities associated with the Schild solution by an over-ring: $\rSchild$, $\uSchild$. Although in this article we focus on this particular circular case, our formalism holds for generic solutions of the Wheeler--Feynman problem. For this reason, in the next few sections we consider a general solution $r_i$ to the Wheeler--Feynman problem, and only restrict to the Schild solution in Section~\ref{sec:results}.

We now substitute Eq.~\eqref{eq:conservativeEOM} into the action~\eqref{eq:expandedAction}, obtaining
\begin{eqnarray}\label{eq:SchildSubtractedAction}
I[A, z_i] &=& - \frac14 \int \dd^4 x \, F_{\mu\nu} F^{\mu\nu} -\sum_i  \int \dd \tau_i \,eQ_i \, A(r_i) \cdot \dot{r}_i \nonumber\\[5pt]
&&-\sum_i \int \dd \tau_i \biggl\{ \tfrac12 m_i \dot z_i^2 - e Q_i z_{i\,\mu} \big( F^{\mu\nu}_\textrm{cons}(r_i) - F^{\mu\nu}(r_i) \big) \dot r_{i \,\nu} \nonumber\\[5pt]
&&\hspace*{20mm} 
+ \frac{e Q_i}{2}z_{i\,\mu}\bigl(z_i^\alpha \partial_\alpha F^{\mu\nu}(r_i)\dot{r}_{i\nu}+F^{\mu\nu}\dot{z}_{i\nu}\bigr)
\\\nonumber&&\hspace*{20mm}+e Q_i\sum_{n=3}^\infty \frac{(n-1)}{n!}
                z_i^{\nu_1}\cdots z_i^{\nu_{n-1}} \dot z_i^{\nu_{n}}
                     \partial_{\nu_1}\cdots \partial_{\nu_{n-2}} F_{\nu_{n-1}\nu_n}(r_i)
\\\nonumber&&\hspace*{20mm}+e Q_i\sum_{n=3}^\infty 
  \frac1{n!} z_i^{\nu_1}\cdots z_i^{\nu_n}
\partial_{\nu_1}\cdots \partial_{\nu_{n-1}} F_{\nu_n\mu}(r_i)\,\dot r_i^\mu 
\biggr\} \,.
\end{eqnarray}
In this action, we are still viewing $A_\mu$ as a variable in the path integral, and $F_{\mu\nu}$ as its field strength. In contrast, $A_\textrm{cons}^\mu$ is a fixed solution with field strength $F^{\mu\nu}_\textrm{cons}(r_i)$.
The appearance of $F^{\mu\nu}_\textrm{cons}(r_i)$ will result in subtractions in certain photon propagators as we will discuss in Section~\ref{sec:unbound}. 
Furthermore, radiation-reaction contributions will also require subtractions on certain photon propagators.
These would correspond to counterterms added to the action~\eqref{eq:SchildSubtractedAction} which we will leave implicit.

We can express interaction vertices in a traditional diagrammatic form --- but of course there is no need to use traditional Feynman diagram techniques to compute the observables.
We will present the full set of interaction vertices and Green's functions below in Section~\ref{sec:frules}. For now, let us focus on the simplest interaction vertex, derived from the first line of \eqref{eq:SchildSubtractedAction}. This vertex is
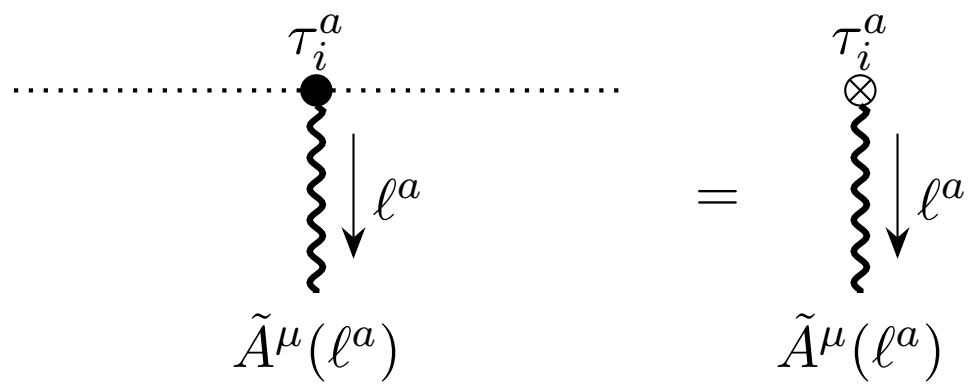
\[\label{eq:verte1}
\begin{tikzpicture}[baseline=(cc.base)]
    \begin{feynman}
        \vertex at (0, 1) (i1);
        \vertex[dot,minimum size=2mm] at (2, 1) (a) {};
        \vertex at (2,-0.7) (b) {\(\Atilde^{\mu}(\ell^a)\)};
        \vertex at (4, 1) (at);
        \vertex at (3,0.2) (cc);
        \node[above] at (a) {$\tau_i^{a}$};
    \diagram*{
        (i1) -- [dotted, thick] (a) -- [dotted, thick] (at),
        (b) -- [photon,very thick,rmomentum'={[arrow distance=7pt]\(\ell^a\)}] (a),
    };
    \end{feynman}
    \end{tikzpicture} \quad 
=\begin{tikzpicture}[baseline=(cc.base)]
    \begin{feynman}
        \vertex at (0, 1) (i1);
        \vertex[crossed dot,minimum size=2mm] at (2, 1) (a) {};
        \vertex at (2,-0.7) (b) {\(\Atilde^{\mu}(\ell^a)\)};
        \vertex at (4, 1) (at);
        \vertex at (3,0.2) (cc);
        \node[above] at (a) {$\tau_i^{a}$};
    \diagram*{
%        (i1) -- [dotted, thick] (a) -- [dotted, thick] (at),
        (b) -- [photon,very thick,rmomentum'={[arrow distance=7pt]\(\ell^a\)}] (a),
    };
    \end{feynman}
    \end{tikzpicture} \quad     
    =-i \, eQ_i 
    %\int_{-\infty}^{\infty} d \tau^a_i
    \,e^{i \ell^a \cdot r_{i}^{a}} \, u^{a\mu}_{i}
\]
where the dotted line represents the unperturbed trajectory of the massive particle.
We leave all integrations implicit.
Here we have used the short-hand notation $u_i^{a\mu}=u_i^{\mu}(\tau^a_i) \coloneq \dot{r}_i^\mu(\tau_i^a)$, and $r_i^{a\mu} = r_i^\mu(\tau_i^a)$.
In our use of Eq.~\eqref{eq:verte1}, $r_i(\tau)$ will be the Schild solution. 
However we could just as easily take $r_i = b_i + u_i \tau$ as in Eq.~\eqref{eq:scatteringTrajectory} to obtain the solution in the scattering context.
This suggests that there should be a quite close correspondence between the raw integrands in the bound and scattering contexts.
Of course, the result of performing the $\tau$ integrals will be different because of the different forms of $r_i(\tau)$, so this correspondence will be obscured at the level of integrated observables.

In the scattering context, we have on-shell conditions $k^2 = 0$ and $\epsilon(k) \cdot k = 0$ for emitted photons.
In addition, we have the on-shell condition $u_i \cdot u_i =1$ for the scalars.
These on-shell conditions carry over to the bound regime, unchanged for the photon, and in the form $u_i^a \cdot u_i^a = 1$ for identically decorated scalar four-velocities.

\section{Power Counting}
\label{sec:powerCounting}

In order to place our perturbative expansion on a solid footing, we perform a dimensional analysis.
For simplicity we focus on a Schild solution where both the masses $m_i$ and the charges $Q_i$ of the particles are comparable. Then the solution is characterised by one length scale $R$ and one mass scale $m$.

Reading off dimensions from the action~\eqref{eq:startingAction} which itself has dimensions of mass $\times$ length, we see that $e$ has dimensions of $\sqrt{\text{mass $\times$ length}}$ while the dimensions of the classical field $A^\mu$ are $\sqrt{\text{mass $/$ length}}$. The dimensionless quantity that we can construct is thus 
\[
\frac{e^2}{m R} \,.
\]
The waveform for photon emission shares the dimensions of the classical field, and so we expect it to be of order $e/R$ at lowest order.
In this paper, we will take corrections to the leading order result to be suppressed by powers of $e^2/(mR)$. The same parameter also governs the expansion of the waveform.  

It is worth comparing to the scattering situation. 
Here the only available length scale is the magnitude of the impact parameter $B = \sqrt{-(b_1 - b_2)^2}$, and therefore the expansion parameter is $e^2 / (mB)$. We will give diagrammatic examples of this perturbative counting below in Section~\ref{sec:unbound}. 
We will refer to the waveform at lowest order in $e$ by `LO' (leading order); to the next-to-leading correction in $e$ by `NLO', and so on.

Our discussion above indicates that it is useful to expand the radiation field as
\newcommand{\Aradpc}[2]{A_{\textrm{rad}}^{#1, (#2)}}
\[
\Arad^\mu(x) = \Aradpc{\mu}{1}(x) + \Aradpc{\mu}{2}(x) + \cdots \,,
\]
where $\Aradpc{\mu}{n}$ is of order $e^{2n - 1}$.
The radiative field strength $F^{\mu\nu}_\textrm{rad}$ follows the same expansion.
Similarly, we can expand the small displacement in a perturbative series:
\[
z_{i}^\mu(\tau)&= \zpc{i}{\mu}{1}(\tau)+ \zpc{i}{\mu}{2}(\tau)   +\cdots \,,
\]
where $\zpc{i}{\mu}{n}$ is of order $e^{2n}$.
Similarly, $\dzpc{i}{\mu}{n}$ and $\ddzpc{i}{\mu}{n}$ are of order $e^{2n}$. 
The displacement $z_i$ grows secularly with time. For small $e$, this growth is small over a single orbital time, and our perturbative expansion is sensible. 

For the circular orbits to which we will be restricting our attention in the present article, the virial theorem tells us that
\[
mv^2 \sim \frac{e^2}{R} \,,
\]
where $v$ is the velocity of one of the charges.
In common with other field theory-based perturbative approaches to bound-state dynamics, the velocity expansion of our waveforms yields terms which are quantitatively comparable to contributions formally at higher orders in $e$. Nevertheless, higher-order velocity corrections at a fixed order in $e$ contain valuable information that can be used to inform inspiral dynamics.

\section{The Perturbative Ingredients}
\label{sec:frules}

We are ultimately interested in computing quantities which can be measured by an observer at large distances --- which we idealize as an observer at infinity.
The primary example of this are the projections of the vector field $A^\mu$, or more particularly its radiative part, onto physical polarizations.
There are other quantities which can only be measured by nearby observers, but which are nonetheless amenable to classical computation.
An example of this is the perturbation $z^\mu$ from the dominant trajectory $r^\mu$.

These observables are in parallel with the waveform and the impulse in the unbound case. 
There we can obtain results for the observables using modern scattering amplitudes techniques.
The same results could also be obtained using a more traditional Feynman diagram expansion.
Our approach here will be to use such a traditional expansion to link to a classical expansion around a bound state solution.
Indeed we will be able to demonstrate a term-by-term equivalence at the level of integrands between the expansions in the two regimes.
This will allow us to present an algorithm taking as input the integrand of a scattering amplitude and yielding corresponding bound-state observables.
We will thereby link the perturbative expansion in powers of $e$ to the perturbative expansion in the unbound case.

The leading terms of the total outgoing field which survive to large distances fall off as $1/|\vec \obsCoord|$ and depend only on the radiation field
\[
\Arad^\mu(x) \coloneq\frac{1}{2}\left( A^\mu_\textrm{{ret}}(x) - A^\mu_\textrm{{adv}}(x) \right) \,.
\] 
These fields are obtained by integrating the retarded or advanced Green's function against the source current,
\begin{eqnarray}\label{eq:totalFieldIntro}
&&A^\mu_\textrm{{ret,\,adv}}(x) = \int d^4 y \, G_\textrm{{ret,\,adv}} (x-y) J^\mu(y) \,,\nonumber\\[5pt]
&&G_\textrm{ret}(x) = \frac{1}{2\pi} \theta(x^0) \delta( x^2) \,,~~~G_\textrm{adv}(x) = \frac{1}{2\pi} \theta(-x^0) \delta(x^2) \,,
\end{eqnarray}
as can also be understood from the KMOC formalism~\cite{Caron-Huot:2023vxl}.
Hence the radiative field may be expressed as
\[\label{eq:onshellfield}
\Arad^\mu(x) =  \int d^4 y\,  G_\textrm{rad}(x-y) J^\mu(y) \,,
\]
where,
\begin{eqnarray}\label{eq:Gradcons}
G_\textrm{cons}(x) &=& \frac{1}{2} \left(G_\textrm{ret}(x)+G_\textrm{adv}(x)\right)= \frac{1}{4\pi} \delta(x^2) \,,\nonumber\\[5pt]
G_{\rm{rad}}(x) &=& \frac{1}{2} \left(G_\textrm{ret}(x)-G_\textrm{adv}(x)\right)=\frac{1}{4\pi} {\rm sgn}(x^0)\delta(x^2)  \,.
\end{eqnarray}

In this paper we are interested in the outgoing waveform (coefficient of $1/r$ as $r = |\mathbf{x}| \to \infty$) at fixed retarded time $u = t-r$.
At nonzero frequency, this waveform is twice the outgoing part of
Eq.~\eqref{eq:onshellfield} due to the normalisation of $G_{\rm{rad}}$ in Eq.~\eqref{eq:Gradcons}. Equivalently, we can compute it directly
using the retarded propagator. This is close in spirit to the on-shell
description of radiation in the KMOC formalism, where virtual and cut
contributions combine to give the retarded prescription for the classical
waveform~\cite{Caron-Huot:2023vxl,Capatti:2024bid}.

We will derive the Feynman diagrammatic representation from the action~\eqref{eq:expandedAction}.
The use of a worldline action makes it straightforward to expand around non-trivial solutions of the equations of motion, such as the Schild solution.
Strictly speaking, we should actually use the equations of motion in order to incorporate the boundary conditions of the Schild solution precisely. 
However, the Schild subtraction in Eq.~\eqref{eq:SchildSubtractedAction} effectively takes this boundary condition into account. 
We can then derive Feynman rules for our perturbative expansion as was shown long ago by Boulware and Brown~\cite{Boulware:1968zz}.
These Feynman rules are equivalent to the ones that would be derived from the equations of motion, up to modifications due to boundary conditions.

In order to derive the following Feynman rules, one substitutes $x_i = r_i + z_i$ into this action and expands relying on the power counting in $e$ as explained in the previous section.
For times small compared to the secular growth time, $z_i$ scales as $e^2$.

The 4-velocity $u_i \coloneq {dr_i}/{d\tau_i}$ is the proper-time derivative of the conservative solution. 
The lowercase index $i=1,2$ on $Q,m,r,u,\tau,$ and $\omega$ label the different worldlines. Photons are labeled by the indices $a,b,\ldots$. When a photon $a$ connects two different worldlines $i$ and $j$, the corresponding proper times at its ends are labeled by $\tau^a_i,\,\tau^a_j$. We also use the shorthand notation $r_i^a \equiv r_i(\tau_i^a)$, and similarly for $u$. In case a photon propagator connects to the \textit{same} worldline at two different proper times, we denote the latter by $\tau_i^{a}$ and $\tau_i^{\bar{a}}$.
The displacements of the $i$th worldline are denoted by $z_i\equiv x_i-r_i$ and their propagators are functions of two proper times $\tau^a_i,\,\tau^b_i$ at their ends.

\newcommand{\action}{I}

We derive the Feynman rules from the action $I[A, z_i]$ of Eq.~\eqref{eq:expandedAction}.
As we make frequent use of retarded propagators, we follow the convention in which time on the worldlines flows from left to right. The retarded $z$ propagator is,
\[
  \begin{tikzpicture}[baseline=(a.base)]
    \node (a) {\(z_i^{a\mu}\)};
    \node (b) [right=3.2cm of a] {\(z_i^{b\nu}\)};
    \draw[thick] (a) -- node[above]{\(\omega_i^{ab}\)} (b);
  \end{tikzpicture} \quad
  = -i\frac{1}{m_i} \frac{\eta^{\mu \nu}}{\left(\omega_i^{ab}+i\epsilon\right)^2}\,.
\]
In our application, the labeling of the proper times at each vertex will be important.
We indicate them by superscripted lower-case initial Latin letters $a, b, c$ etc, and abbreviate $z_i^\mu(\tau^a) = z_i^{a\mu}$.
In turn, we label the propagator frequencies $\omega_i^{ab}$ by the labels of the two vertices it connects.
We defer the integral over the frequencies to the end of the calculation.
As we will discuss in Section~\ref{sec:unbound}, we wish to focus on properties of the integrand.

The usual momentum-space Feynman photon propagator is, of course,
\[
  \begin{tikzpicture}[baseline=(a.base)]
  \begin{feynman}
    \node (a) {\(\Atilde^{\mu}(\ell^a)\)};
    \node (b) [right=3.5cm of a] {\(\Atilde^{\nu}(\ell^a)\)};
    \diagram*{
        (a) -- [photon, very thick, momentum'={[arrow distance=7pt]\(\ell^a\)}, edge label=\(F\)] (b);
    };
    \end{feynman}
    \end{tikzpicture} \quad
  = \frac{-i}{(\ell^a)^2 + i \varepsilon} \eta^{\mu\nu}\,.
\]
We wrote the propagator in Feynman gauge for simplicity, and included the symbol $F$ in the diagram to emphasise that the propagator satisfies Feynman boundary conditions.
We also label the photon momentum with the index $a$ associated with the proper time of its emission, and write $\tilde A^\mu$ for the Fourier transform of $A^\mu$.
Of course, the full photon propagator is diagonal in the momentum $\ell^a$; this forces the vertex label $a$ at both ends of the photon line to be the same (with a caveat we will discuss below). The momentum space representation of the retarded Green's function (or propagator) \eqref{eq:totalFieldIntro} is 
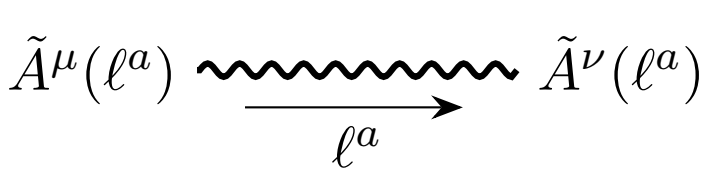
\[
  \begin{tikzpicture}[baseline=(a.base)]
  \begin{feynman}
    \node (a) {\(\Atilde^{\mu}(\ell^a)\)};
    \node (b) [right=3.5cm of a] {\(\Atilde^{\nu}(\ell^a)\)};
    \diagram*{
        (a) -- [photon, very thick, momentum'={[arrow distance=7pt]\(\ell^a\)}] (b);
    };
    \end{feynman}
    \end{tikzpicture} \quad
  = \frac{-i}{(\ell^a)^2 + i \, \ell^0 \epsilon} \eta^{\mu\nu}\,.
\]
It will further be convenient to identify the part of the photon propagator corresponding to the radiative part of the field, namely the momentum space representation of $\tilde G_\text{rad}$ from Eq.~\eqref{eq:Gradcons}, 
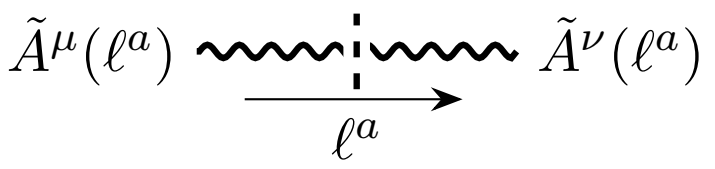
\[
\label{eq:cutp}
  \begin{tikzpicture}[baseline=(a.base)]
  \begin{feynman}
    \node (a) {\(\Atilde^{\mu}(\ell^a)\)};
    \node (b) [right=3.5cm of a] {\(\Atilde^{\nu}(\ell^a)\)};
    \diagram*{
        (a) -- [photon, onshell=0.5, very thick, momentum'={[arrow distance=9pt]\(\ell^a\)}] (b);
    };
    \end{feynman}
    \end{tikzpicture} \quad
  = - \frac{1}2 \hat{\delta}(\ell^2) \sgn(\ell^0) \eta^{\mu\nu} \,.
\]
We will refer to this as a ``subtracted'' photon propagator, for reasons which will become clear shortly. As with the frequencies $\omega^{ab}$, we defer the integrals over $\ell$ to the end of the computation --- as usual, these integrals will be regulated dimensionally.

The explicit vertices are written with the convention of outgoing momenta (and frequencies) as indicated in the diagrams.
Ordinarily, we would integrate over $\tau^a_i$ immediately; but in order to focus on the properties of the integrand, we defer the integrations over all proper-time variables.
Accordingly, the Feynman diagrams in this article have a slightly different meaning from the usual diagrams used in the scattering regime.

Our Feynman vertices also descend directly from the action \eqref{eq:SchildSubtractedAction}. In momentum space, we note that every derivative $\partial_\rho$ acting on $A^\mu$ becomes a product with momentum $i\ell_\rho\widetilde{A}^\mu$, while $\dot{z}^\mu_i$ becomes $i\omega z^\mu_i$. Additionally, the Fourier transform to momentum space adds a factor $e^{i \left[\ell \cdot r+(\sum\omega)\tau\right]}$ where $\ell$ is the momentum of the outgoing photon, and the sum is over all frequencies $\omega$ in the vertex. 
The first vertex in our Feynman rules is the emission vertex from the unperturbed trajectory, already given in \eqref{eq:verte1}. This vertex originates from the last term on the first line of \eqref{eq:SchildSubtractedAction}. The next vertex describes the interaction between a photon and a displacement: 
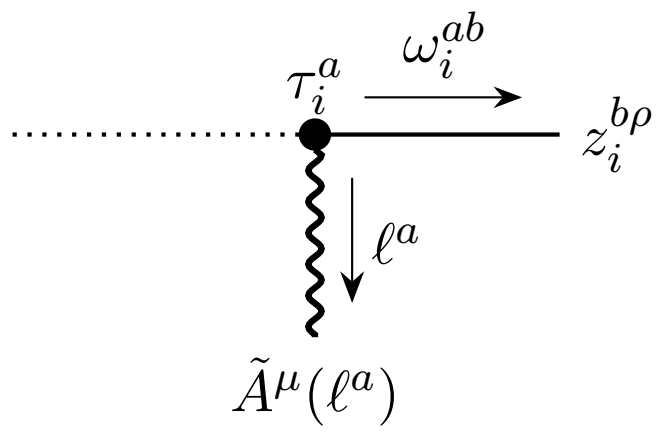
\begin{eqnarray}
\begin{tikzpicture}[baseline=(cc.base)]
    \begin{feynman}
        \vertex at (0, 1) (i1) ;
        \vertex[dot,minimum size=2mm] at (2, 1) (a) {};
        \vertex at (2,-0.7) (b) {\(\Atilde^{\mu}(\ell^a)\)};
        \vertex at (4, 1) (at) {\(z_{i}^{b\rho}\)};
        \vertex at (3,-0.2) (cc);
        \node[above] at (a) {$\tau_i^{a}$};
    \diagram*{
        (i1) -- [dotted, thick] (a) -- [thick,momentum={[arrow distance=7pt]\(\omega^{ab}_i\)}] (at),
        (b) -- [photon,very thick,rmomentum'={[arrow distance=7pt]\(\ell^a\)}] (a),
    };
    \end{feynman}
    \end{tikzpicture}&=&
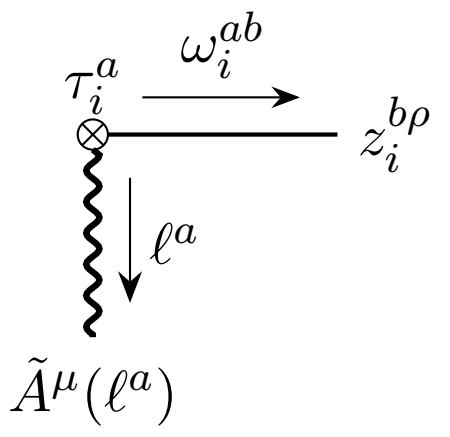
\begin{tikzpicture}[baseline=(cc.base)]
    \begin{feynman}
        \vertex at (0, 1) (i1) ;
        \vertex[crossed dot,minimum size=2mm] at (2, 1) (a) {};
        \vertex at (2,-0.7) (b) {\(\Atilde^{\mu}(\ell^a)\)};
        \vertex at (4, 1) (at) {\(z_{i}^{b\rho}\)};
        \vertex at (3,-0.2) (cc);
        \node[above] at (a) {$\tau_i^{a}$};
    \diagram*{
        (a) -- [thick,momentum={[arrow distance=7pt]\(\omega^{ab}_i\)}] (at),
        (b) -- [photon,very thick,rmomentum'={[arrow distance=7pt]\(\ell^a\)}] (a),
    };
    \end{feynman}
    \end{tikzpicture} =    
    eQ_i 
    %\int_{-\infty}^{\infty} d \tau^a_i
    \,e^{i \left[\ell^a \cdot r_{i}^a+\omega_i^{ab}\tau^a_i\right]} \left( \ell^{a\rho} u^{a\mu}_{i}+\omega^{ab}_i\eta^{\rho\mu}\right)\nonumber\\[5pt]
    &=&eQ_i 
    %\int_{-\infty}^{\infty} d \tau^a_i
    \,e^{i \left[\ell^a \cdot r_{i}^a+\omega_i^{ab}\tau^a_i\right]} \ell^{a[\rho} \eta^{\beta]\mu}u_{i\beta}^{a}\,.
\label{eq:SingleZVertex}
\end{eqnarray}
This vertex originates from the last term on the second line of \eqref{eq:SchildSubtractedAction}. In the second form of the Feynman rule, the crossed dot indicates the presence of the background field, which is shown explicitly in the first (left) version of the vertex. The second equality for this vertex makes use of integration by parts (IBP) on the integrated trajectory. In this vertex and all other vertices, we can take the momenta $\ell^a$ and frequencies $\omega^{ab}_i$ to be complex.

It will be convenient to use the crossed-dot depiction of the vertex in additional vertices. 
The third vertex we need describes interactions between displacements and photons:
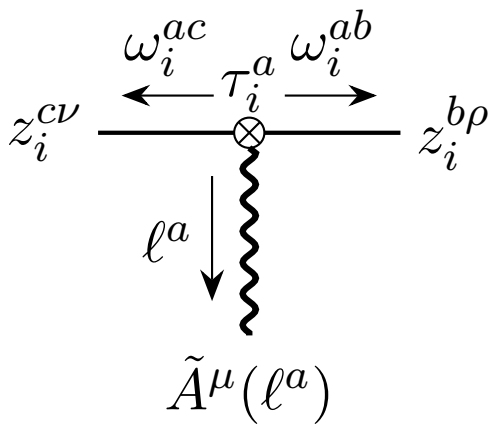
\[
\label{eq:twozvertex}
    \begin{tikzpicture}[baseline=(cc.base)]
    \begin{feynman}
        \vertex at (1, 1) (i1) ;
        \node   at ($(i1)+( -10pt, 0pt )$) {\(z_{i}^{c\nu}\)};
        \vertex[crossed dot,minimum size=2mm] at (2, 1) (a) {};
        \vertex at (2,-0.7) (b) {\(\tilde{A}^{\mu}(\ell^a)\)};
        \vertex at (3, 1) (at);
        \node   at ($(at)+( 10pt, 0pt )$) {\(z_{i}^{b\rho}\)};
        \vertex at (2,-0.2) (cc);
        \node[above] at (a) {$\tau_i^{a}$};
    \diagram*{
        (i1) -- [thick,rmomentum={[arrow distance=7pt]\(\omega^{ac}_i\)}] (a) -- [thick,momentum={[arrow distance=7pt]\(\omega^{ab}_i\)}] (at),
        (b) -- [photon,very thick,rmomentum={[arrow distance=7pt]\(\ell^a\)}] (a),

    };
    \end{feynman}
    \end{tikzpicture}=i \,eQ_i 
    %\int_{-\infty}^{\infty} d \tau^a_i
    \,e^{i \left[\ell^a \cdot r_{i}^{a}+(\omega_i^{ac}+\omega_i^{ab})\tau_i^a\right]}\,\left[\ell^{a\nu} \ell^{a\rho} u^{a\mu}_{i}+\omega_i^{ac} \ell^{a\rho}\eta^{\mu\nu}+\omega_i^{ab} \ell^{a\nu}\eta^{\rho\mu}\right]\,.\\
\]
This vertex originates from the last term on the third line of \eqref{eq:SchildSubtractedAction}.
Evidently, the final two sums of \eqref{eq:SchildSubtractedAction} give rise to vertices with a higher multiplicity of $z$'s, though we will not need them at the order explored in this paper. Finally, the $F^{\mu\nu}_{\rm cons}$ term on the second line of \eqref{eq:SchildSubtractedAction} leads to a tadpole contribution to $z^\mu$, in the form
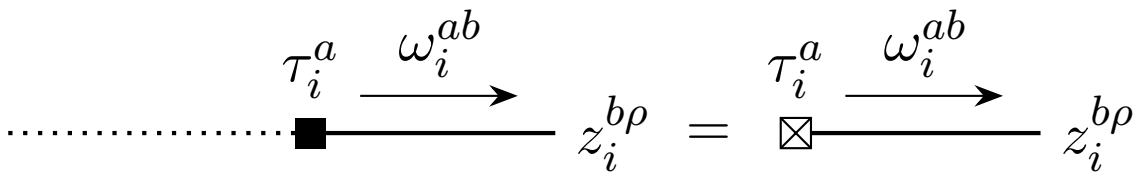
\begin{equation}
\label{eq:tadpole}
\begin{aligned}
    \begin{tikzpicture}[baseline={([yshift=-0.5ex]a.center)}]
        \begin{feynman}
            \vertex at (0,0) (i1);
            \vertex[shape=rectangle, fill=black, minimum size=2mm,
                    inner sep=0pt, outer sep=0pt] at (2,0) (a) {};
            \vertex at (4,0) (at) {\(z_i^{b\rho}\)};
            \node[above] at (a.north) {$\tau_i^a$};
            \diagram*{
                (i1) -- [dotted, thick] (a)
                     -- [thick, momentum={[arrow distance=7pt]\(\omega_i^{ab}\)}] (at),
            };
        \end{feynman}
    \end{tikzpicture}
    &=
    \begin{tikzpicture}[baseline={([yshift=-0.5ex]a.center)}]
        \begin{feynman}
            \vertex[shape=rectangle, draw=black, fill=white, minimum size=2mm,
                    inner sep=0pt, outer sep=0pt] at (0,0) (a) {};
            \vertex at (2,0) (at) {\(z_i^{b\rho}\)};
            \node[above] at (a.north) {$\tau_i^a$};
            \diagram*{
                (a) -- [thick, momentum={[arrow distance=7pt]\(\omega_i^{ab}\)}] (at),
            };
            \draw (a.north west) -- (a.south east)
                  (a.south west) -- (a.north east);
        \end{feynman}
    \end{tikzpicture}
    \\[5pt]
    &= i e Q_i e^{i \omega_i^{ab} \tau_i^a}
       F_{\text{cons}}^{\rho\nu}\left(r_i^a\right)u_{i\nu}^a\,.
\end{aligned}
\end{equation}
We denote the tadpole vertex by a square, to differentiate it from other vertices involving a dynamical photon. Because $F^{\mu\nu}_{\rm cons}$ is itself $\mathcal{O}(e)$, the tadpole is of order $\mathcal{O}(e^2)$. (The counterterm, like $F^{\mu\nu}_{\rm cons}$, is proportional to the charge of the other particle.) Although in QED, the usual power counting is one power of momentum per (trivalent) vertex, in our application higher powers of momenta arise due to the Taylor expansion of the action.
Of course, the same Taylor expansion leads to an infinite set of vertices which we will not describe here.
Each of these vertices is of order $e$, but will require diagrams involving more and more vertices in order to obtain a diagram describing classical observables such as the waveform.
Hence, these additional vertices effectively contribute only at higher order.

\section{Connecting Unbound to Bound Observables}
\label{sec:unbound}

We can use the Feynman rules developed in Section~\ref{sec:frules} to compute observables with different background solutions $r_i(\tau)$.
In particular, we could choose $r_i^\mu(\tau) = b_i^\mu + u_i^\mu \tau$ to be an unbound solution, as in Eq.~\eqref{eq:scatteringTrajectory}.
This would yield observables in a scattering encounter.
We could instead choose $r_i(\tau)$ to be the Schild solution, leading to observables in a circular binary.
(We could alternatively use standard amplitudes techniques to derive the observables for a scattering encounter, and the integrands would be the same as those derived from the Feynman rules of Section~\ref{sec:frules}.)

The rules, before integration over proper times $\tau$, or equivalently over the frequencies $\omega$, are in fact \emph{identical} until we substitute a specific form of $r_i(\tau)$.
The substitution of specific forms for $r_i$, and the integrations, will lead of course to different results.
This suggests that we should be able to map the integrands for unbound observables to those for bound-state observables.
This is in fact what we will find in selected examples below.
We conjecture that this mapping holds in general.

In the following subsections, we discuss the simplest, and next-to-simplest, contribution to the waveform.
The simplest contribution is of order $e$, and the next-to-simplest is of order $e^3$.
In the unbound case, the simplest contribution actually vanishes after integration because of on-shell conditions.
In the bound case, this simplest contribution is non-vanishing and gives the leading-order waveform.
The next-to-simplest contribution accordingly gives the leading-order waveform in the unbound case, and the next-to-leading contribution in the bound case.

In the subsections below, we use the notation $\integrand^\mu_n$ for contributions to the integrands of waveforms at order $e^n$.
The integrated unbound waveform we denote by $\waveA_n$ and the integrated bound state waveform will be denoted by $\waveB_n$. Both $\waveA_n$ and $\waveB_n$ are \textit{amputated}, in the sense that they are defined without the propagator of the emitted photon.

\subsection{The simplest terms in the waveform}

Using the Feynman rules from Section~\ref{sec:frules}, the simplest contribution to the integrand of the waveform is given by single photon emission from one of the worldlines, with no other photon exchanges, 
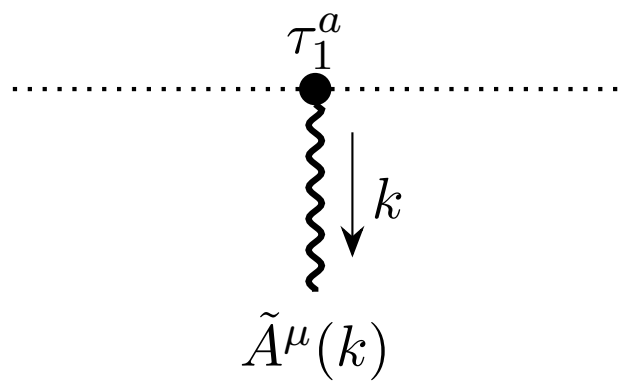
\[\label{eq:simplest}
i\integrand_1^\mu=\begin{tikzpicture}[baseline=(cc.base)]
    \begin{feynman}
        \vertex at (0, 1) (i1);
        \vertex[dot,minimum size=2mm] at (2, 1) (a) {};
        \vertex at (2,-0.7) (b) {\(\tilde{A}^{\mu}(k)\)};
        \vertex at (4, 1) (at);
        \vertex at (3,-0.2) (cc);
        \node[above] at (a) {$\tau_1^{a}$};
    \diagram*{
        (i1) -- [dotted, thick] (a) -- [dotted, thick] (at),
        (b) -- [photon,very thick,rmomentum'={[arrow distance=7pt]\(k\)}] (a),
    };
    \end{feynman}
    \end{tikzpicture} \quad &=-i \, eQ_1 
    %\int_{-\infty}^{\infty} d \tau^a_1
    \,\exp \big(i \, k \cdot r_{1}^{a} (\tau_1^a)\big) \, u^{a\mu}_{1}(\tau_1^a) \,.
\]
There is a similar contribution for emission from particle 2, which can be obtained by replacing the particle label $1 \to 2$.

We can obtain the waveform itself by performing the proper time integral.
Here we have to specialise $r_1^a$ and its derivative $u_1^a$ to a particular choice of background.
For the scattering problem, we take $r_1^a$ as given by Eq.~\eqref{eq:scatteringVacuum}.
The waveform is then,
\[\label{eq:0orderScattering}
i \waveA_1^\mu(k) &= -i \, eQ_1 \, u^{\mu}_{1} \int_{-\infty}^{\infty} d \tau^a_1
    \,\exp \big(i \, k \cdot( b_1 + u_1 \tau_1^a )\big) \\
    &= -i \, eQ_1 \, u^{\mu}_{1} \, e^{i \, k \cdot b_1} \, \hat\delta(k \cdot u_1) \,.
\]
Notice that the four-velocity $u_1$ here is constant, and therefore we do not include a label $a$.
We contract with the photon's polarisation vector, and rewrite the four-velocity in terms of the particle momentum to find,
\[
i \polvector^*(k) \cdot \waveA_1(k) = -2i \, e Q_1 \, p_1 \cdot \polvector^*(k) \, e^{i \, k \cdot b_1} \hat\delta(2 \,k \cdot p_1) \,. 
\]
We recognise this as the three-point scattering amplitude (times a phase reflecting the impact parameter), with the energy-momentum conserving delta function implicit. 
The $\hat\delta$ factor is the on-shell condition for one of the massive legs, which also forces the waveform to vanish for real momenta.

In the bound case, we instead choose the Schild trajectories~\eqref{eq:schildStarter}, and again perform the proper time integral from $-\infty$ to $\infty$.
The result is (c.f. \cite{Goldberger:2017vcg}),
\[
i \waveB^\mu &= -i \, eQ_1 \, \int_{-\infty}^{\infty} d \tau^a_1
    \,\exp \big(i \, k \cdot \rSchild_1(\tau_1^a)\big) \, \uSchild^{\mu}_{1} (\tau_1^a) \\
&= - i \, e Q_1 \, f^\mu(k) \,.
\]
The function $f^\mu$ is a non-trivial function, which we calculate in Section~\ref{sec:results}. 

This is the simplest example of a waveform, and it demonstrates that integration leads to very different results from the same integrand.
It may be worth remarking that the electromagnetic radiation from the circular bound system is nothing but synchrotron radiation.

On-shell gauge invariance in Eq.~\eqref{eq:0orderScattering} is trivial, because $k \cdot \waveA_1$ is proportional to
\[
k \cdot u_1 \, \hat{\delta}(k \cdot u_1) = 0 \,.
\]
Although $f^\mu(k)$ is non-trivial, on-shell gauge invariance is nevertheless straightforward because $k \cdot \waveB$ is proportional to,
\[
i \int_{-\infty}^{\infty} d \tau^a_1
    \,\exp \big(i \, k \cdot \rSchild_1(\tau_1^a)\big) \, k \cdot \uSchild_{1} (\tau_1^a) &= 
    \int_{-\infty}^{\infty} d \tau^a_1
    \,\frac{d}{d\tau_1} \exp \big(i \, k \cdot \rSchild_1(\tau_1^a)\big)
    = 0 \,.
\]
Similar identities will play a role in the following subsections.

\subsection{The next-to-simplest waveform}

We next examine contributions to the waveform involving a single photon exchanged between the two massive charged particles. At the same $\mathcal{O}(e^3)$, we also have a diagram with a tadpole for the displacement $z^\mu$. The overall integrand $H^\mu_3$ is then
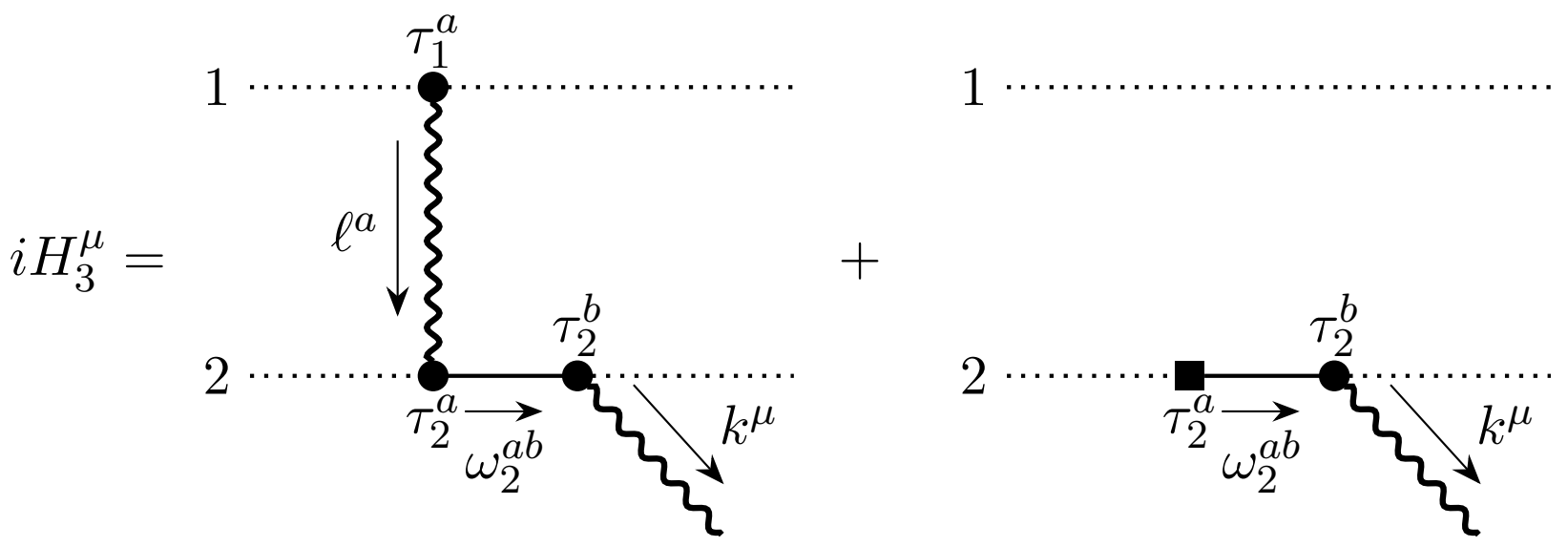
\begin{figure}[h]
    \centering
    
    \raisebox{0.7cm}{\(i \integrand^\mu_3=\)}% \hspace{0.75em}
    \tikzset{outgoing photon label/.style={xshift=4pt, yshift=-5pt}}
    \begin{tikzpicture}[baseline=(l2.center)]
        \begin{feynman}
            \vertex at (0,1) (l1) {\(1\)};
            \vertex[dot, minimum size=2mm] at (1.5,1) (t1a) {};
            \vertex at (4,1) (r1);

            \vertex at (0,-1) (l2) {\(2\)};
            \vertex[dot, minimum size=2mm] at (1.5,-1) (t2a) {};
            \vertex[dot, minimum size=2mm] at (2.5,-1) (t2b) {};
            \vertex at (4,-1) (r2);
            \vertex at (3.5,-2.1) (k);

            \node[above] at (t1a) {$\tau_1^{a}$};
            \node[below] at (t2a) {$\tau_2^{a}$};
            \node[above] at (t2b) {$\tau_2^{b}$};

            \diagram*{
                (l1) -- [dotted, thick] (t1a)
                     -- [dotted, thick] (r1),
                (l2) -- [dotted, thick] (t2a),
                (t2b) -- [dotted, thick] (r2),
                (t1a) -- [
                    photon, very thick,
                    momentum'={[arrow distance=7pt]\(\ell^{a}\)}
                ] (t2a),
                (t2a) -- [
                    plain, thick,
                    momentum'={[arrow distance=7pt]\(\omega_2^{ab}\)}
                ] (t2b),
                (k) -- [
    photon, very thick,
    rmomentum'={[arrow distance=7pt,
                 label style=outgoing photon label]\(k^{\mu}\)}
] (t2b),
            };
        \end{feynman}
    \end{tikzpicture}%
    \hspace{0.75em}%
\raisebox{0.7cm}{\(+\)}%
\hspace{0.75em}%
\tikzset{outgoing photon label/.style={xshift=4pt, yshift=-5pt}}
    \begin{tikzpicture}[baseline=(l2.center)]
        \begin{feynman}
            \vertex at (0,1) (l1) {\(1\)};
            \vertex at (4,1) (r1);

            \vertex at (0,-1) (l2) {\(2\)};
            \vertex[shape=rectangle, fill=black, minimum size=2mm,
                    inner sep=0pt, outer sep=0pt] at (1.5,-1) (t2a) {};
            \vertex[dot, minimum size=2mm] at (2.5,-1) (t2b) {};
            \vertex at (4,-1) (r2);
            \vertex at (3.5,-2.1) (k);

            \node[below] at (t2a) {$\tau_2^{a}$};
            \node[above] at (t2b) {$\tau_2^{b}$};
            \diagram*{
                (l1) -- [dotted, thick] (r1),
                (l2) -- [dotted, thick] (t2a),
                (t2b) -- [dotted, thick] (r2),
                (t2a) -- [
                    plain, thick,
                    momentum'={[arrow distance=7pt]\(\omega_2^{ab}\)}
                ] (t2b),
                (k) -- [
    photon, very thick,
    rmomentum'={[arrow distance=7pt,
                 label style=outgoing photon label]\(k^{\mu}\)}
] (t2b),
            };
        \end{feynman}
    \end{tikzpicture}
        \hspace{0.75em}%
\caption{The $Q_1 Q_2^2$ contribution to the integrand at $\mathcal{O}(e^3)$.}\label{fig:placeholder}
\end{figure}

\begin{figure}[h]
    \centering
    
    \raisebox{0.7cm}{\(i \integrand^\mu_3=\)}% \hspace{0.75em}
    \tikzset{outgoing photon label/.style={xshift=4pt, yshift=-5pt}}
    \begin{tikzpicture}[baseline=(l2.center)]
        \begin{feynman}
            \vertex at (0,1) (l1) {\(1\)};
            \vertex[dot, minimum size=2mm] at (1.5,1) (t1a) {};
            \vertex at (4,1) (r1);

            \vertex at (0,-1) (l2) {\(2\)};
            \vertex[dot, minimum size=2mm] at (1.5,-1) (t2a) {};
            \vertex[dot, minimum size=2mm] at (2.5,-1) (t2b) {};
            \vertex at (4,-1) (r2);
            \vertex at (3.5,-2.1) (k);

            \node[above] at (t1a) {$\tau_1^{a}$};
            \node[below] at (t2a) {$\tau_2^{a}$};
            \node[above] at (t2b) {$\tau_2^{b}$};

            \diagram*{
                (l1) -- [dotted, thick] (t1a)
                     -- [dotted, thick] (r1),
                (l2) -- [dotted, thick] (t2a),
                (t2b) -- [dotted, thick] (r2),
                (t1a) -- [
    photon, onshell=0.5, very thick,
    momentum'={[arrow distance=7pt]\(\ell^{a}\)}
] (t2a),
                (t2a) -- [
                    plain, thick,
                    momentum'={[arrow distance=7pt]\(\omega_2^{ab}\)}
                ] (t2b),
                (k) -- [
    photon, very thick,
    rmomentum'={[arrow distance=7pt,
                 label style=outgoing photon label]\(k^{\mu}\)}
] (t2b),
            };
        \end{feynman}
    \end{tikzpicture}%
        \hspace{0.75em}%
\caption{Summing the tadpole is equivalent to using a subtracted propagator.}\label{fig:subtract}
\end{figure}
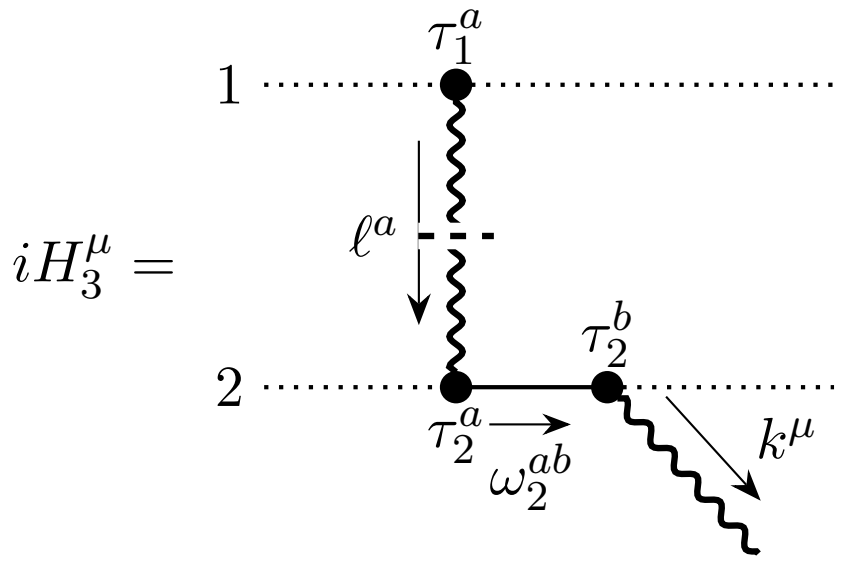

We remind the reader that the temporal flow along the worldlines is from left to right in our conventions. By our Feynman rules, the tadpole diagram is equal to the photon exchange diagram before it, albeit with a negative sign and a conservative, rather than a retarded propagator. Adding the two is then equivalent to one photon exchange diagram with a radiative, or subtracted, propagator, as shown in Fig.~\ref{fig:subtract}.
\FloatBarrier
Applying the Feynman rules yields,
\begin{eqnarray}\label{eq:hNLO}
\begin{aligned}
i H_3 \cdot \varepsilon^*(k)= & -\frac{e^3 Q_1 Q_2^2}{2 m_2} \frac{\operatorname{sgn}\left(\ell^{a 0}\right) \hat{\delta}\bigl(\left(\ell^a\right)^2\bigr)}{\left(\omega_2^{a b}+i \epsilon\right)^2} e^{i \ell^a \cdot\left(r_1^a-r_2^a\right)+i \omega_2^{a b}\left(\tau_2^a-\tau_2^b\right)+i k \cdot r_2^b} \\
& \times\left[\left(u_1^a \cdot u_2^a\right) \ell^{a \mu}-\omega_2^{a b} u_1^{a \mu}\right]\left[k_\mu\bigl(u_2^b \cdot \varepsilon^*\bigr)-\omega_2^{a b} \varepsilon_\mu^*\right] .
\end{aligned}
\end{eqnarray}
Notice that the $u$s and $r$s in this expression carry labels $a$ and $b$. 
These labels are significant as they indicate the different proper times at which the positions and four-velocities are evaluated.
In the scattering context, the distinction between different $u$s would disappear because they are constant, but this is not true in the bound binary context.
In the bound case, we obtain the corresponding waveform by evaluating Eq.~\eqref{eq:hNLO} with $r$ taken to be $\rSchild$ and $u$ taken to be $\uSchild$.

Alternatively, using standard vacuum Feynman rules in scalar QED, we could compute the amplitude and the associated integrand of the waveform using KMOC. 
We would then obtain $H_3$ with $r_i$ specialised to the scattering trajectory, and with a retarded photon propagator rather than a subtracted one.
We see again that there is a close connection between the integrands for the two scenarios even though --- as we shall see later --- the results of the final integrals will be quite different.

\subsection{Radiation reaction and the NLO waveform}
\label{sec:nlo}

There is another contribution to the waveform at order $e^3$ involving no photon exchange between the two lines:

\begin{figure}[h]
    \centering
\begin{tikzpicture}[baseline=(cc.base)]
    \begin{feynman}
        \vertex at (0, -1) (i1) ;
        \vertex[dot,minimum size=2mm, label= below:$\tau_2^a$] at (1.5, -1) (u) {};
         \vertex[dot,minimum size=2mm, label= below:$\tau_2^{\bar{a}}$] at (0.5, -1) (u0) {};
         \vertex[dot,minimum size=2mm, label= below:$\tau_2^b$] at (2.5, -1) (sf) {};
        \vertex at (4, -1) (at);
        \vertex at (0, 1) (i1d) ;
         \vertex at (4, 1) (atd);
         \vertex at (3.5, 0.2) (m);
        \node   at ($(m)+( 12pt, -5pt )$) {$k$};
        \vertex at (3,0) (cc);
    \diagram*{
        (i1) -- [dotted, thick] (u),
        (m) -- [photon, very thick] (sf),
        (sf) -- [dotted, thick] (at),
        (sf)-- [thick,rmomentum'={[arrow distance=7pt]\(\omega_2^{ab}\)}](u),
        (i1d)-- [dotted, thick] (atd),
        (u) -- [photon,bend right= 50, very thick, rmomentum'={[arrow shorten=0.1,arrow distance=7pt]\(\ell^{\bar{a}a}\)}] (u0),
    };
    \end{feynman}
    
    \end{tikzpicture}
    \caption{The radiation-reaction contribution proportional to $Q^3_2$ to the $\mathcal{O}(e^3)$ integrand.}
    \label{fig:self-energy}
\end{figure}
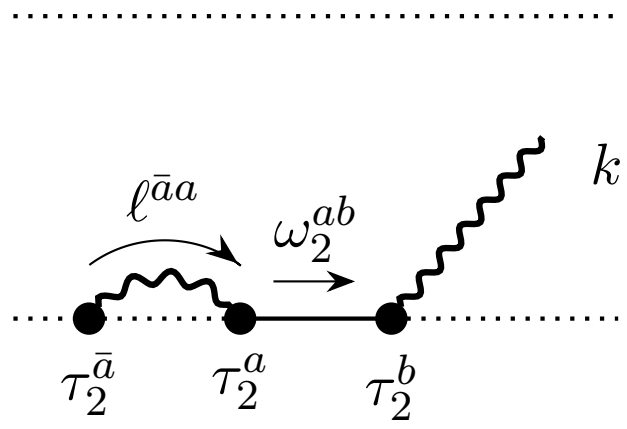

Its contribution is proportional to $Q_2^3$, and is structurally very similar to the NLO waveform $H_3$ of Eq.~\eqref{eq:hNLO}.
In this particular case, we can obtain the result very simply by substituting $\tau_1^a \to \tau_2^{\bar{a}}$, and hence also $r_1^a \to r_2^{\bar{a}}$, $u_1^a \to u_2^{\bar{a}}$,
relabeling $\ell^a \to \ell^{\bar{a}a}$, and further replacing $Q_1 \to Q_2$. 
In this case, we cannot identify the proper times at the two ends of the photon line (this is the caveat mentioned in Section~\ref{sec:frules}).
We then find,
\begin{eqnarray}\label{eq:finalH3}
i \integrand_3' \cdot \polvector^*(k)&=& -\frac{i e^3 Q_2^3}{m_2\left(\ell^2+i \ell^0\epsilon\right) }\frac{1}{\left(\omega_2^{a b}+i \epsilon\right)^2} e^{i \ell^{\bar{a} a} \cdot\left(r_2^{\bar{a}}-r_2^{a}\right)+i \omega_2^{a b}\left(\tau_2^{a}-\tau_2^b\right)+i k \cdot r_2^b} \nonumber\\[5pt]
&& \times\left[\left(u_2^{\bar{a}} \cdot u_2^{a}\right) \ell^{\bar{a}a \mu}-\omega_2^{a b} u_2^{\bar{a} \mu}\right]\left[k_\mu\bigl(u_2^b \cdot \varepsilon^*\bigr)-\omega_2^{a b} \varepsilon_\mu^*\right] .
\end{eqnarray}
Here we used the shorthand $\ell^{\bar{a}a \mu}\rightarrow\ell$ in the propagator.
This contribution is formally divergent; it can be regulated similarly to the derivation of the Abraham--Lorentz--Dirac (ALD) force \cite{Dirac1938,Detweiler:2002mi}. This would amount to performing a subtraction on the (self-energy) photon propagator.

While this contribution vanishes in the scattering context, the next order contribution (with an additional exchanged photon) would not.
The self-energy-like part of the diagram should be thought of as corresponding to an internal bubble in the KMOC amplitude.
The emergence of a radiation-reaction contribution in the NLO waveform matches the emergence of such a contribution in the NNLO impulse.
In the scattering context, this happens at order $e^5$ in the waveform; in the bound state context, at order $e^3$.

\subsection{The general A2B algorithm}\label{sec:algorithm}

Based on the observations above, we propose an algorithm for obtaining the bound-state integrand from ordinary scattering computations.
Here, we restrict to massive-scalar electrodynamics.
We expect the generalisation to the electrodynamics of spinning particles to be straightforward, but the generalisation to gravity will require non-trivial further development.
We begin with a slightly modified integrand of a scattering amplitude.
That is, we compute a slightly modified~\cite{Caron-Huot:2023vxl} KMOC integrand (taking care to include the KMOC cut) deferring all loop and Fourier integrals but including the $i\epsilon$ prescriptions and on-shell delta functions.
We assume an analytic expression for the integrand.
The algorithm will produce integrands suitable for a generic background.
We are agnostic about the technology used to compute the integrand; one could use BCFW combined with generalised unitarity, Feynman diagrams, Berends-Giele recursion, analytic reconstructions from numerical unitarity~\cite{Abreu:2020xvt}, or even unitarity alone proceeding from maximal cuts to next-to-maximal and so on.

The on-shell delta functions in the KMOC expression should be rewritten as $\tau$ integrals over exponential factors, with again the $\tau$ integrals deferred.

We proceed as follows. 
First note that the scattering amplitudes we have in mind have two distinct massive lines running through them.

\begin{enumerate}
\item In computing the amplitude, we keep track of subamplitudes
(or correspondingly vertices in a diagrammatic version), assigning
each a label, for example with initial Latin letters $a$, $b$, $c$ 
and so on. We will sometimes call these labels `decorations'.
We write all momenta in terms of the incoming momenta of the scalar particles, and the independent 
photon propagator momenta.
We keep separate impact parameters for the two
massive lines. \label{algo:decorations}

\item We rewrite the incoming scalar momenta in terms of the four velocities $u_i$ associated with the massive lines. 
These velocities inherit the labels of the subamplitudes in which they appear.
(We could obtain the standard scattering domain amplitude by removing all the decorations.)
\label{algo:pToU}

\item The integrands of self-energy contributions at the same order must be included. 
Each self-energy bubble necessarily contains one photon line and one scalar line. 
(When all contributions are summed over, the propagator(s) of the scalar line(s) will be eikonalised and averaged over the Feynman and anti-Feynman prescriptions.)
Care must be taken to incorporate the correct on-shell delta function in these cases: there will be two contributions to the argument of the delta function with opposite signs, and different decorations of the scalar momenta. 
We will give an example in Section~\ref{sec:Q_2^3}.
\label{algo:self-energy}

\item Rewrite the on-shell delta functions as proper-time integrals over phase factors.
In performing this rewriting, choose the sign of the argument of the delta function so
that the phase is positive for an outgoing momentum, and negative for an incoming momentum.
The integrals will be left implicit, so this effectively 
trades the delta functions for phase factors.  The
proper times are indexed by the massive line
originating the on-shell delta function,
and also inherit the decorations of the momenta
they are multiplying. 
This effectively introduces multiple proper-time variables for each massive line.
At this stage, we introduce a product of additional delta functions in the proper times, each having the form $\delta(\tau_i^a - \tau_i^b)$, to set these different variables to be equal. 
The integrations over all these additional proper-time variables will also be left implicit.
\label{ProperTimeIntroduction}

\item Each massive propagator is associated to a frequency variable $\omega$ which is labeled by the decorations at the two ends of the massive propagator. 
Rewrite the proper-time delta functions introduced at
step \ref{ProperTimeIntroduction} as integrals of phase factors depending on the appropriate frequency variable, for example $\omega^{ab}$.  
This may require rearranging arguments of the introduced delta functions so that the decorations of the proper times match those of the frequencies.  
Leave the integrals implicit, so this again amounts to trading delta functions for phase factors.\label{FrequencyIntroduction}

\item Each photon propagator connects either the two massive lines, or one of the massive lines to itself.  
We identify the labels at the two ends in all photon momenta in the former case.
Each photon momentum then carries one or two of these labels.
We do not identify the labels of proper times or velocities.
\label{algo:onePhotonDecoration}

\item In practice, all denominators arising from massive propagators will have the form $u_i \cdot \Sigma \, \ell$ (or such quantities squared)~\cite{Capatti:2024bid}. 
Replace the dot products in the denominators by frequency variables (or sums thereof) with the replacement rules deduced from (scattering-domain) IBP identities on the proper time integrals, or equivalently, identities that would be imposed by the proper-time integrations.
\label{DenominatorReplacement}

\item In the decorated scattering amplitude, we identify terms
which require a Schild subtraction.  We will explain a mechanical 
procedure for doing this below.  It may be possible
to construct a companion set of subtracting amplitudes which are built incorporating one or more basic Schild subtraction amplitudes and/or ALD subtraction amplitudes. 
We leave an exploration of the connection between the types of subtraction to future investigation.
\label{also:subtractions}

\end{enumerate}

\def\Ord{\mathcal{O}}
In the worldline theory, the propagators
of exchanged photons connecting a background
to a vertex with a single worldline 
deflection~\eqref{eq:SingleZVertex} require a 
subtraction~\eqref{eq:cutp} as reflected in the last term on the second line
of Eq.~\eqref{eq:SchildSubtractedAction}.  Other photon propagators
should not be subtracted.  We first
take the maximal cut of the amplitude with respect
to all photons, and consider separately
each `half-amplitude', corresponding to a two-scalar
multiple-photon amplitude.  Our goal is to identify the terms in
each decorated half-amplitude which correspond to single-$z$ vertices.

The key observation is that we can link the worldline and
[classical limit of the] ordinary form of the
multiple-emission
scattering amplitude by focusing on the double poles in the eikonalized
matter propagators.  
Suppose we now consider $n+1$ emissions, corresponding to one of the amplitudes
needed for the $\Ord(e^{2n+1})$ waveform.
Because the denominators are linearly dependent,
extracting the coefficients by multivariate
complex analysis is somewhat subtle.  It would
involve multiplying
the amplitude by a lone power of all eikonalized propagators,
and taking residues with respect to $n$ of them at a time.  
The linear dependence means that the choice of contour is subtle, because the integrand can develop singularities on the contour.  
A careful analysis could partly sidestep these subtleties by making use of multivariate partial fractioning on the integrand, using packages 
such as \textsf{MultivariateApart}~\cite{Heller:2021qkz} 
and \textsf{LinApart3}~\cite{Fekeshazy:2026nbn}.
We instead rely on the existence of a special form for the integrand
guaranteed by the worldline formulation, and extract
coefficients of products of squared denominators using
an ansatz.
We solve for the ansatz coefficients.
We can then rely on the fact that
the classical limit will be expressible purely in terms of
on-shell field strengths $F$, one for each external photon.
This representation allows us to add terms to each of the ansatz coefficients in order to complete it into a factor built out of world-line vertices.
We can then identify the terms, and factors within them, that have the form of single-$z$ vertices.
When we sew the photon multi-emission amplitudes together to obtain the integrand of the waveform, photons attached to one single-$z$ vertex are candidates for subtraction.
We give a nontrivial example of this procedure in 
Appendix~\ref{app:MappingProcedure}.

We will discuss the comparison between the results of this algorithm and direct worldline computation in the following subsections at the next-to-leading order in the bound domain.
We have checked that the algorithm agrees with direct computation at one higher order, but we have no proof that this extends to all orders.

\subsection{Example: $Q_1 Q_2^2$ contributions}

Let's see how the algorithm described in the 
previous subsection works at the $Q_1Q_2^2$ order. 
We start with the integrand of the scattering-domain
amplitude, expressed therefore in terms of
straight-line particle trajectories~\eqref{eq:scatteringVacuum}.
Assuming the notation of Eq.~(5.46) of ref.~\cite{Kosower:2018adc}, use the energy-momentum conserving delta function to eliminate $\bar w_2=k-\bar w_1$; we also substitute $b\rightarrow b_1-b_2$, $p_i\rightarrow m_i u_i$ (no sum on $i$), and replace $\polvector$ by $\polvector^*$ to match the conventions of the present paper.
We also replace $\bar w_1\rightarrow \ell^a$ and retain $k=\ell^b$, according to step~\ref{algo:decorations} of the procedure described in the previous subsection.  
We decorate with the interaction labels which would have been present if the amplitude had been constructed according to steps~\ref{algo:decorations} and~\ref{algo:pToU}.
We then obtain,
\begin{equation}
\begin{aligned}
 &-\frac{i e^3 Q_1Q_2^2}{m_2}
 e^{i\ell^a \cdot b_1}
 e^{i(-\ell^a \cdot b_2 + k \cdot b_2)} 
 \delh(\ell^a\cdot u^a_1) \, 
 \delh(-\ell^a \cdot u_2^a+k\cdot u_2^b)  \\
 &\qquad\times \frac{1}{\ell^a \cdot \ell^a} 
 \biggl(u_1^a \cdot \polvector^* 
 -\frac{1}{k \cdot u_2^b}
 \left[ u_1^a\cdot u_2^a \,\ell^a\cdot \polvector^* 
 +  u_1^a \cdot k\,u_2^b \cdot \polvector^*\right] \\
 &\qquad\qquad\qquad +\frac{1}{(k \cdot u_2^b)^2}
 u_1^a \cdot u_2^a\,\ell^a\cdot k \,
 u_2^b \cdot \polvector^*  \biggr)\,.
\end{aligned}
\end{equation}
If we drop the interaction labels $a$ and $b$ on the $u_i$, we recover Eq.~(5.46) of ref.~\cite{Kosower:2018adc} up to the harmless momentum-relabellings above.

Following step~\ref{ProperTimeIntroduction}, 
we rewrite the two delta functions in
terms of integrals
over newly introduced proper times $\tau_{1,2}$,
with the indices linked to the massive lines.
Each proper time variable will show up
in the phase factor as $\ell^x u_i \tau_i$;
it acquires the label of the corresponding
momentum in the term, as does the four-velocity.
We then identify $b_i+u_i^x \tau_i^x$ as $r_i^x$,
with the $r$s inheriting the corresponding label.
We leave the $d\tau_i$ integrals implicit. 
This yields,
\begin{equation}
    \begin{aligned}
   &-\frac{i e^3 Q_1Q_2^2}{m_2}\, 
   e^{i\ell^a \cdot r^a_1}
   e^{i(-\ell^a \cdot r^a_2 + k \cdot r^b_2)}
   \delta(\tau_2^a-\tau_2^b)\\
 &\qquad\times \frac{1}{\ell^a \cdot \ell^a} 
 \biggl(u_1^a \cdot \polvector^* 
 -\frac{1}{k \cdot u_2^b}
 \left[ u_1^a\cdot u_2^a \,\ell^a\cdot \polvector^* 
 +  u_1^a \cdot k\,u_2^b \cdot \polvector^*\right] \\
 &\qquad\qquad\qquad +\frac{1}{(k \cdot u_2^b)^2}
 u_1^a \cdot u_2^a\,\ell^a\cdot k \,
 u_2^b \cdot \polvector^*  \biggr)\,.
\end{aligned}
\end{equation}
Following step~\ref{FrequencyIntroduction},
we express the delta function of the 
proper-time difference as an
integral over a frequency $\omega_2^{ab}$,
which we can think of as a world-line momentum.
We leave the integral over $\dhat\omega_2^{ab}$
implicit.  Our integrand then becomes,
\begin{equation}
    \begin{aligned}
    &-\frac{i e^3 Q_1Q_2^2}{m_2}\,
    e^{i(\ell^a \cdot r^a_1)}
    e^{i(-\ell^a \cdot r^a_2 +k \cdot r^b_2)} e^{i \omega_2^{ab}(\tau_2 ^a -\tau_2^b)}
 \\
 &\qquad\times \frac{1}{\ell^a \cdot \ell^a} 
 \biggl(u_1^a \cdot \polvector^* 
 -\frac{1}{k \cdot u_2^b}
 \left[ u_1^a\cdot u_2^a \,\ell^a\cdot \polvector^* 
 +  u_1^a \cdot k\,u_2^b \cdot \polvector^*\right] \\
 &\qquad\qquad\qquad +\frac{1}{(k \cdot u_2^b)^2}
 u_1^a \cdot u_2^a\,\ell^a\cdot k \,
 u_2^b \cdot \polvector^*  \biggr)\,.
\end{aligned}
\end{equation}

Notice that were we to perform the $\tau_2^b$ integral, for example, we would obtain a delta
function setting $k\cdot u_2^b= \omega_2 ^{ab}$.
Alternatively, we can view this as an IBP relation (it will no longer hold as-is in the bound-state
setting).
We must use this relation to replace
denominators arising from massive propagators
by functions of frequency variables,
as in step~\ref{DenominatorReplacement}.
We may then write our integrand as,
\begin{equation}
\begin{aligned}
    &-\frac{i e^3 Q_1Q_2^2}{m_2}\,
    e^{i(\ell^a \cdot r^a_1)}
    e^{i(-\ell^a \cdot r^a_2 +k \cdot r^b_2)} e^{i \omega_2^{ab}(\tau_2 ^a -\tau_2^b)}
 \\
 &\qquad\times \frac{1}{\ell^a \cdot \ell^a} 
 \biggl(u_1^a \cdot \polvector^* 
 -\frac{1}{\omega_2 ^{ab}}
 \left[ u_1^a\cdot u_2^a \,\ell^a\cdot \polvector^* 
 +  u_1^a \cdot k\,u_2^b \cdot \polvector^*\right] \\
 &\qquad\qquad\qquad +\frac{1}{(\omega_2 ^{ab})^2}
 u_1^a \cdot u_2^a\,\ell^a\cdot k \,
 u_2^b \cdot \polvector^*  \biggr)\,.
\end{aligned}
\end{equation}
Taking the $(\omega_2 ^{ab})^{-2}$ factor out of the brackets, restoring its retarded $i\epsilon$ prescription, and rearranging, we have
\begin{equation}\label{eq:resexchNLO}
\begin{aligned}
    &-\frac{ie^3 Q_1 Q_2^2}{ m_2}\frac{1}{\ell^a\cdot\ell^a} \frac{1}{\left(\omega_2^{a b}+i \epsilon\right)^2} e^{i \ell^a \cdot\left(r_1^a-r_2^a\right)+i \omega_2^{a b}\left(\tau_2^a-\tau_2^b\right)+i k \cdot r_2^b} \\
& \times\left[\left(u_1^a \cdot u_2^a\right) \ell^{a \mu}-\omega_2^{a b} u_1^{a \mu}\right]\left[k_\mu\bigl(u_2^b \cdot \varepsilon^*\bigr)-\omega_2^{a b} \varepsilon_\mu^*\right]\,.
\end{aligned}
\end{equation}
In the unbound regime, the photon propagator is retarded. In the bound regime the propagator is initially retarded, but together with the tadpole it becomes subtracted, or radiative. This gives exactly \eqref{eq:hNLO}.

\subsection{Example: $Q_2^3$ contributions}
\label{sec:Q_2^3}

\begin{figure}[h]
\centering
\begin{tikzpicture}[scale=1.0, 
                baseline={([yshift=-\the\dimexpr\fontdimen22\textfont2\relax]
                    current bounding box.center)},
        ] 
\begin{feynman}
  \vertex (i) {};
  \vertex [right=1.5 of i] (c);
  \node [below=0.4 of c] {$c$};
  \vertex [right=2 of c] (ab);
  \node [below=0.3 of ab] {$ab$};
  \vertex [right=1.5 of ab] (f) {};
  \vertex [above=1.5 of f] (k) {};
  \vertex [above=2 of i] (i2) {};
  \vertex [above=2 of f] (f2) {};
  \draw [fermion, thick, momentum'={[arrow shorten=0.3,arrow distance=8pt]\(p_2\)}] (i) -- (c);
  \draw [fermion, thick] (c) -- (ab);
  \draw [fermion, thick, momentum'={[arrow shorten=0.3,arrow distance=8pt]\(p_2'\)}] (ab) -- (f);
  \draw [fermion, thick] (f2) -- (i2);
  \diagram* {
    (c) -- [photon, thick, bend left=90, momentum={[arrow shorten=0.3,arrow distance=8pt]\(\ell^{ca}\)}] (ab);
    (ab) -- [photon, thick, momentum={[arrow shorten=0.3,arrow distance=8pt]\(k^b\)}] (k);
  };
%  \filldraw [color=black] (c) circle [radius=2pt];
  \filldraw [color=white] (ab) circle [radius=7pt];
  \filldraw [fill=allOrderBlue] (ab) circle [radius=7pt];
\end{feynman}
\end{tikzpicture}
\caption{One contribution to the self-energy. The arrows on the solid lines indicate the charge flow on the scalar lines. Other orderings should be included to obtain the full self-energy.}
\label{fig:self-energyAmps}
\end{figure}
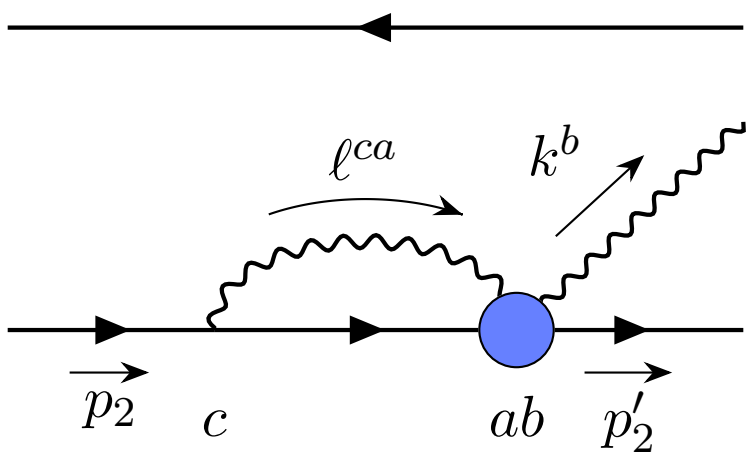

In this section, we examine how to obtain the radiation-reaction contribution previously discussed in Section~\ref{sec:nlo}.
Contributions of this type, as shown in Fig.~\ref{fig:self-energyAmps} are ordinarily dropped (in scattering amplitudes) because they are disconnected; they also involve an outgoing photon which (again in scattering) carries zero four-momentum.
We can nonetheless write down an expression for the integrand using standard techniques as decorated in steps~\ref{algo:decorations} and~\ref{algo:pToU} of the A2B algorithm. 

In the following we omit contributions associated with the upper line in Fig.~\ref{fig:self-energyAmps}. 
Our starting point is one of two contributions to 
the integrand computed following steps~\ref{algo:decorations} and~\ref{algo:pToU} of our algorithm. 
The KMOC integrand contains an on-shell delta function of the form $\delh\bigl(\left(p_2^{\prime}\right)^2-m_2^2\bigr)$.
The first term contains terms quadratic in the messenger momenta $\ell^{ca}$ and $k^b$ which do not survive the classical limit, and which we can ignore. 
The terms linear in the messenger momenta involve dot products with the scalar momentum, which we must decorate with the vertex, or basic three-point amplitude, labels as we build up this contribution.
This gives rise to a delta function of the form $\delh(k^b\cdot u_2^b + \ell^{ca} \cdot u_2^c- \ell^{ca} \cdot u_2^a)$.
The decorated contribution shown in Fig.~\ref{fig:self-energyAmps} is then,
\begin{equation}
\begin{aligned}
 &\frac{ie^3Q_2^3}{m_2}
 e^{ik^b \cdot b_2}
 \delh(k^b\cdot u_2^b + \ell^{ca} \cdot u_2^c- \ell^{ca} \cdot u_2^a) \frac{i}{u_2^c \cdot \ell^{ca} - i \epsilon} \\
 &\qquad\times \frac{1}{\ell^{ca} \cdot \ell^{ca}} 
 \biggl(u_2^c \cdot \polvector^* 
 -\frac{1}{\ell^{ca} \cdot u_2^a}
 \left[ u_2^a\cdot u_2^c \,\ell^{ca}\cdot \polvector^* 
 +  u_2^c \cdot k^b\,u_2^b \cdot \polvector^*\right] \\
 &\qquad\qquad\qquad +\frac{1}{(\ell^{ca} \cdot u_2^a)^2}
 u_2^a \cdot u_2^c\,\ell^{ca}\cdot k^b \,
 u_2^b \cdot \polvector^*  \biggr)\,.
\end{aligned}
\end{equation}
The second contribution is related to this one, up to overall sign, by taking $u_2^x \rightarrow - u_2^x$. 

Again, following step~\ref{ProperTimeIntroduction}, 
we rewrite the delta function in terms of an integral over a newly introduced proper time $\tau_{2}$, corresponding to the second massive line.
The proper time variable will show up in the phase factor as $\ell^x u_2 \tau_2$;
each instance acquires the label of the corresponding momentum in the term, as does the four-velocity.
We introduce delta functions to set the differently-decorated proper time variables to be equal.
We then identify $b_2+u_2^x \tau_2^x$ as $r_2^x$,
with the $r$s inheriting the corresponding decoration.
We leave the $d\tau_2^x$ integrals implicit. 
This yields, 
\begin{equation}
\begin{aligned}
 &\frac{ie^3Q_2^3}{m_2} e^{i k^b \cdot r_2^b} e^{i \ell^{ca} \cdot (r_2^c-r_2^a)} 
 \delta(\tau_2^a - \tau_2^b) \delta(\tau_2^a - \tau_2^c) \frac{i}{u_2^c \cdot \ell^{ca} - i \epsilon} \\
 &\qquad\times \frac{1}{\ell^{ca} \cdot \ell^{ca}} 
 \biggl(u_2^c \cdot \polvector^* 
 -\frac{1}{\ell^{ca} \cdot u_2^a}
 \left[ u_2^a\cdot u_2^c \,\ell^{ca}\cdot \polvector^* 
 +  u_2^c \cdot k^b\,u_2^b \cdot \polvector^*\right] \\
 &\qquad\qquad\qquad +\frac{1}{(\ell^{ca} \cdot u_2^a)^2}
 u_2^a \cdot u_2^c\,\ell^{ca}\cdot k^b \,
 u_2^b \cdot \polvector^*  \biggr)\,.
\end{aligned}
\end{equation}
Following step~\ref{FrequencyIntroduction},
we express the delta functions of the proper-time differences as integrals over two frequencies $\omega_2^{ab}$ and $\omega_2^{ac}$, which we can think of as world-line momenta.
We leave the integrals over $\dhat\omega_2^{ab}$ and $\dhat\omega_2^{ac}$ implicit.  Our integrand then becomes,
\begin{equation}
\begin{aligned}
 &\frac{ie^3Q_2^3}{m_2} e^{i k^b \cdot r_2^b} e^{i \ell^{ca} \cdot (r_2^c-r_2^a)} 
 e^{i \omega_2^{ab} (\tau_2^a - \tau_2^b)} e^{i \omega_2^{ac} (\tau_2^a - \tau_2^c)} \frac{i}{u_2^c \cdot \ell^{ca} - i \epsilon} \\
 &\qquad\times \frac{1}{\ell^{ca} \cdot \ell^{ca}} 
 \biggl(u_2^c \cdot \polvector^* 
 -\frac{1}{\ell^{ca} \cdot u_2^a}
 \left[ u_2^a\cdot u_2^c \,\ell^{ca}\cdot \polvector^* 
 +  u_2^c \cdot k^b\,u_2^b \cdot \polvector^*\right] \\
 &\qquad\qquad\qquad +\frac{1}{(\ell^{ca} \cdot u_2^a)^2}
 u_2^a \cdot u_2^c\,\ell^{ca}\cdot k^b \,
 u_2^b \cdot \polvector^*  \biggr)\,.
\end{aligned}
\end{equation}
Were we to perform the $\tau_2^c$ integral, for example, we would obtain a delta
function setting $\ell^{ca}\cdot u_2^c= \omega_2 ^{ac}$.
Alternatively, we can view this as an IBP relation (it will no longer hold as-is in the bound-state
setting).  We must use this relation to replace
denominators arising from massive propagators
by functions of frequency variables,
as in step~\ref{DenominatorReplacement}.

We may then write our integrand as,
\begin{equation}
\begin{aligned}
 &\frac{ie^3Q_2^3}{m_2} e^{i k^b \cdot r_2^b} e^{i \ell^{ca} \cdot (r_2^c-r_2^a)} 
 e^{i \omega_2^{ab} (\tau_2^a - \tau_2^b)} e^{i \omega_2^{ac} (\tau_2^a - \tau_2^c)} \frac{i}{\omega_2^{ac} - i \epsilon} \\
 &\qquad\times \frac{1}{\ell^{ca} \cdot \ell^{ca}} 
 \biggl(u_2^c \cdot \polvector^* 
 -\frac{1}{\omega_2^{ab} + \omega_2^{ac}}
 \left[ u_2^a\cdot u_2^c \,\ell^{ca}\cdot \polvector^* 
 +  u_2^c \cdot k^b\,u_2^b \cdot \polvector^*\right] \\
 &\qquad\qquad\qquad +\frac{1}{(\omega_2^{ab} + \omega_2^{ac})^2}
 u_2^a \cdot u_2^c\,\ell^{ca}\cdot k^b \,
 u_2^b \cdot \polvector^*  \biggr)\,.
\end{aligned}
\end{equation}
In order to match the worldline form exactly, we should combine both contributions. When we do so, we find an integrand of the form,
\[
\frac{iX}{\omega_2^{ac} - i \epsilon} - \frac{iX}{\omega_2^{ac} + i \epsilon} = -X\delh(\omega_2^{ac}) \,.
\]
Performing the $\omega_2^{ac}$ integral, we then obtain 
\[
 &\frac{-ie^3Q_2^3}{m_2} e^{i k^b \cdot r_2^b} e^{i \ell^{ca} \cdot (r_2^c-r_2^a)} 
 e^{i \omega_2^{ab} (\tau_2^a - \tau_2^b)}  \\
 &\qquad\times \frac{1}{\ell^{ca} \cdot \ell^{ca}} 
 \biggl(u_2^c \cdot \polvector^* 
 -\frac{1}{\omega_2^{ab}}
 \left[ u_2^a\cdot u_2^c \,\ell^{ca}\cdot \polvector^* 
 +  u_2^c \cdot k^b\,u_2^b \cdot \polvector^*\right] \\
 &\qquad\qquad\qquad +\frac{1}{(\omega_2^{ab})^2}
 u_2^a \cdot u_2^c\,\ell^{ca}\cdot k^b \,
 u_2^b \cdot \polvector^*  \biggr)\,,
\]
Taking the $(\omega_2 ^{ab})^{-2}$ factor out of the brackets, restoring its retarded $i\epsilon$ prescription, and rearranging, we have
\begin{eqnarray}
 &&-\frac{i e^3 Q_2^3}{m_2\left(\ell^2+i \ell^0\epsilon\right) }\frac{1}{\left(\omega_2^{a b}+i \epsilon\right)^2} e^{i \ell^{c a} \cdot\left(r_2^{c}-r_2^{a}\right)+i \omega_2^{a b}\left(\tau_2^{a}-\tau_2^b\right)+i k \cdot r_2^b} \nonumber\\[5pt]
&& \times\left[\left(u_2^{c} \cdot u_2^{a}\right) \ell^{ca \mu}-\omega_2^{a b} u_2^{c \mu}\right]\left[k_\mu\bigl(u_2^b \cdot \varepsilon^*\bigr)-\omega_2^{a b} \varepsilon_\mu^*\right] .
\end{eqnarray}
Here we used the shorthand $\ell^{ca \mu}\rightarrow\ell$ in the propagator. This matches Eq.~\eqref{eq:finalH3}, with $c\rightarrow\bar a$.

\section{Quasi-Circular Orbits: Explicit Results}
\label{sec:results}

In this section, we focus on the particular choice $r\rightarrow \rSchild$, namely, we take our baseline to be the Schild solution~\cite{Schild:1963} discussed earlier in Section~\ref{sec:pertSetup}. 
The circular trajectories of the Schild solution are given in \eqref{eq:schildStarter}. 
The corresponding four-velocities are given by,
\begin{eqnarray}\label{eq:schild2}
\uSchild^\mu_i(\tau) \coloneq  \frac{d \rSchild^\mu_i}{d\tau} =\gamma_i\left(1,\mp v_i\sin\left[\gamma_i \Omega\tau\right],\pm v_i\cos \left[\gamma_i \Omega\tau\right],0\right)\,.
\end{eqnarray}
The quantities $\gamma_i$ are the usual Lorentz boost factors $\gamma_i^2 = 1/(1 - v_i^2)$, $\Omega$ is a common angular frequency, and $v_i$ are coordinate velocities of each body.
This solution has a conserved center-of-momentum energy given by
\begin{eqnarray}\label{eq:schen}
E=m_1\gamma^{-1}_1+m_2\gamma^{-1}_2\,.
\end{eqnarray}
The angular frequency $\Omega$ and coordinate velocities can be viewed as functions of the energy $E$. 
See appendix~\ref{app:Schild} for more details.
In what follows, we compute the far-field radiation emitted by the perturbed Schild solution and the associated energy loss up to NLO. As a consistency check, we show that our results reduce to the standard non-relativistic expressions in the non-relativistic limit.
\subsection{Far-field radiation from the Schild solution: LO result} \label{farfieldradLO}

As a starting point, we compute the emitted radiation in the far field from
the Schild solution, corresponding to the limit $r,t\to\infty$ with fixed
retarded time $u\equiv t-r$. For convenience, we compute the components
$\polvector^{\pm}_{\mu} A_{\rm rad}^{\mu,(1)}$ in the circular basis about
the orbital $z$ axis, where
$\polvector^\mu_{\pm}=(0,1,\pm i,0)/\sqrt{2}$.
We first calculate the radiated field in momentum space from diagram~\eqref{eq:simplest} and its sibling diagram with the photon emitted from particle 2.
Attaching the radiative propagator~\eqref{eq:cutp} to the
integrated diagrams gives
\begin{equation}\label{eq:Schildrad}
    {\polvector}_\mu^{\pm}\tilde{A}_{\rm rad}^{\mu,(1)}(k)
    =-\frac{1}{2}\,\hat{\delta}(k^2)\operatorname{sgn}(k^0)\,
      i{\polvector}_{\pm}\cdot\left[\mathcal{B}^{1,0}+\mathcal{B}^{0,1}\right]\,.
\end{equation}
As discussed in Sec.~\ref{sec:frules}, at nonzero frequency the emitted waveform is twice
the outgoing part of this field. After Fourier transformation we therefore
retain the term depending on $u=t-r$, multiply it by two, and omit the
incoming term depending on $t+r$. We denote the resulting waveform by
$A_{\rm out}^{\mu,(1)}$.

We write $\mathcal{B}^{p,q}$ for the contribution to the integrated amputated bound waveform
proportional to $e^{p+q}Q_1^p Q_2^q$, including these factors. Thus $\mathcal{B}_1=\mathcal{B}^{1,0}+\mathcal{B}^{0,1}$ and
$\mathcal{B}_3=\mathcal{B}^{1,2}+\mathcal{B}^{2,1}+\mathcal{B}^{3,0}+\mathcal{B}^{0,3}$. Explicitly,
\begin{eqnarray}\label{eq:B1def}
   &&i{\polvector}_{\pm}\cdot \mathcal{B}^{1,0} =-ie Q_{1}\int d\tau^a_1 \, e^{ik \cdot \rSchild_1^a}\,\uSchild_{1}^{a}\cdot {\polvector}_{\pm}\,,
\end{eqnarray}
and similarly for $i {\polvector}_{\pm}\cdot \mathcal{B}^{0,1}$. The computation of $i {\polvector}_{\pm}\cdot \mathcal{B}^{1,0}$ is straightforward using the Jacobi-Anger expansion:
\begin{eqnarray}
    e^{-iz\cos\theta}=\sum_{n=-\infty}^{\infty}\left(-i\right)^{n}J_{n}\left(z\right)e^{-in\theta}
\end{eqnarray}
where $J_n$ denotes the Bessel function of the first kind. The result is
\begin{eqnarray}\label{eq:B10res}
   &&i{\polvector}_{\pm}\cdot \mathcal{B}^{1,0} =\mp\frac{eQ_{1}v_{1}}{\sqrt{2}}\sum_{n=-\infty}^{\infty}\hat{\delta}\left(k^{0}-n\Omega\right)\left(-i\right)^{n\pm1}e^{i(n\pm1)\varphi^{k}}J_{n\pm1}\left(k_{\perp} R_{1}\right)\,.
\end{eqnarray}
where $R_1\equiv \Omega^{-1} v_1,\,\hat{t}^\mu\equiv(1,0,0,0),\,k_\perp\equiv \sqrt{(k^{1})^2+(k^{2})^2}$, and $\varphi^k$ is the azimuthal angle of $\vec{k}$. For future reference, we also define
\begin{eqnarray}\label{eq:C10def}
   &&C^{1,0} =e Q_{1}\int d\tau^a_1 \, e^{ik \cdot \rSchild_1^a}=\frac{e Q_1}{\gamma_1} \sum_{n=-\infty}^{\infty} \hat{\delta}\left(k^{0}-n \Omega\right)(-\mathrm{i})^n e^{\mathrm{i} n \varphi^k} J_n\left(k_{\perp} R_1\right)\,,
\end{eqnarray}
and a similarly defined $C^{0,1}$. To obtain $A_{\rm rad}^{\mu,(1)}$ in
coordinate space, we substitute Eq.~\eqref{eq:B10res} and its particle 2
counterpart into Eq.~\eqref{eq:Schildrad} and take an inverse Fourier transform,
\begin{equation} \label{eq:iF}
    {\polvector}_\mu^{\pm} {A}_{\rm{rad}}^{\mu,(1)} \left(x\right)=\int \hat{d}^4k\,e^{-i(k^{0} t-\vec{k}\cdot\vec{x})}\,{\polvector}_\mu^{\pm} \tilde{A}_{\rm{rad}}^{\mu,(1)} \left(k\right)\,\,.
\end{equation}
This is made simpler by using the angular stationary-phase formula in the
far-field limit,
\begin{equation}
    e^{i\vec{k}\cdot\vec{x}}
    \sim\frac{1}{2\pi i k r}\left[   e^{ikr}\hat{\delta}\left(\hat{\vec{k}}-\hat{\vec{x}}\right)
      -e^{-ikr}\hat{\delta}\left(\hat{\vec{k}}+\hat{\vec{x}}\right)
      \right]\,.
\end{equation}
Here $k=|\vec{k}|$, and the hatted delta functions on the sphere are normalized with respect to $d\Omega_{\hat{\vec{k}}}/(2\pi)^2$. Integrating and then discarding the incoming term, which depends on $t+r$, and doubling the
outgoing term, which depends on $u=t-r$, we obtain, 
\begin{equation} \label{eq:Schildradsimp1}
    \left[{\polvector}_\mu^{\pm} {A}_{\rm{out}}^{\mu,(1)} \left(x\right)\right]_{\rm particle\, 1} = -\frac{eQ_1 v_1}{4\pi\sqrt{2}r} \,\sum_{n=-\infty}^{\infty}\left(-i\right)^{n} e^{i (n\pm 1)\varphi}  J_{n\pm 1}\left(n v_1 \sin\theta\right)e^{-i n \Omega u}\,\,.
\end{equation}
The field emitted from particle 2 is obtained by relabeling $Q_1,\,v_1 \to Q_2,\,v_2$,
shifting $\varphi\to\varphi+\pi$, and including an overall minus sign from the reversal
of the spatial velocity. Overall, defining $\bar{\varphi}^{j}\equiv \varphi + \pi \delta_{j,2}$
and $s_j=(-1)^{j+1}$ with $j = 1,2$, we write
\begin{equation} \label{eq:Schildradsimp}
    {\polvector}_\mu^{\pm} {A}_{\rm{out}}^{\mu,(1)} \left(x\right) = -\sum_{j=1,2}\,\frac{eQ_j s_jv_j}{4\pi\sqrt{2}r} \,\sum_{n=-\infty}^{\infty}\left(-i\right)^{n} e^{i (n\pm 1)\bar{\varphi}^j}  J_{n\pm 1}\left(n v_j \sin\theta\right)e^{-i n \Omega u}\,\,.
\end{equation}
To obtain the gauge-invariant helicity components of the emitted waveform, we now project onto the circular basis transverse to $\hat{\vec{x}}$.
Writing $\epsilon_\lambda^\mu=(0,\hat{\boldsymbol\theta}
+i\lambda\hat{\boldsymbol\varphi})/\sqrt{2}$ with $\lambda=\pm1$, the
helicity components are $\epsilon_\lambda^*\cdot A_{\rm out}^{(1)}$.
For the planar current considered here the Lorenz-gauge potential has
$A_{\rm out}^{z,(1)}=0$, so the projection follows directly from~\eqref{eq:Schildradsimp}:
\begin{equation} \label{eq:Schildhelicity}
    \epsilon_\lambda^*\cdot A_{\rm out}^{(1)}
    =\frac12\sum_{s=\pm1}(\cos\theta-\lambda s)e^{-is\varphi}\,
     {\polvector}_s\cdot A_{\rm out}^{(1)}\,.
\end{equation}

\subsection{LO energy flux}
As explained earlier, the physical trajectory of the two charges can be thought of as hopping from one Schild solution to another, corresponding to an evolution of the energy $E$. As is well known, the energy flux at infinity is the integral of the Poynting vector, that is,
\begin{align} \label{eq:energyloss}
    P^{\infty} = \int d\Omega\,r^{2}\,|\vec{E}^{\perp}_{\rm{out}}|^{2}\,,
\end{align}
where $\vec{E}^{\perp}_{\rm{out}}$ is the transverse electric field in the far zone. By transverse, we mean that it is perpendicular to $\vec{r}$.
The leading-order contribution to the loss comes from the electric field generated by a Schild trajectory, namely derived from the potential~\eqref{eq:Schildradsimp}. Using the helicity components in Eq.~\eqref{eq:Schildhelicity}, the LO contribution is
\begin{align}
    \vec{E}^{\perp,(1)}_{\rm out}
    &= \partial_u \sum_{\lambda=\pm 1}
       \vec{\epsilon}_{\lambda}
       \left(\epsilon_{\lambda}^{*}\cdot A_{\rm out}^{(1)}\right)
       \nonumber\\
    &= \partial_u\left[
       \left(\polvector^{-}\cdot A_{\rm out}^{(1)}\right)
       \bar{\vec{\polvector}}^{+}
       + \left(\polvector^{+}\cdot A_{\rm out}^{(1)}\right)
       \bar{\vec{\polvector}}^{-}
       \right],
    \label{eq:E1}
\end{align}
where $\vec{\epsilon}_{\lambda}$ is the spatial part of $\epsilon_{\lambda}^{\mu}$ and
$\bar{\vec{\polvector}}^{\pm}
= \vec{\polvector}^{\pm}
-\hat{\vec{x}}\bigl(\hat{\vec{x}}\cdot\vec{\polvector}^{\pm}\bigr)$
are the projections of the orbital polarization vectors onto the plane perpendicular to $\hat{\vec{x}}$. 
This gives rise to
 \begin{align}\label{eq:LOflux}
     P^{\infty}_{\rm LO}=\frac{e^2\Omega^{2}}{4\pi}\sum_{n=1}^{\infty} n^{2}\sum_{j,k}f^n_{jk}\,\mathcal{S}^{njk} \,,
 \end{align}
where $f^n_{jk}\equiv Q_{j}Q_{k}\left(-1\right)^{n\left(\delta_{j,2}+\delta_{2,k}\right)}$, and
   \begin{align} \mathcal{S}^{njk}\equiv\int_{0}^{\pi}d\theta\,\sin\theta\,\left[\cot^{2}\theta J_{n}\left(nv_{j}\sin\theta\right)J_{n}\left(nv_{k}\sin\theta\right)+v_{j}v_{k}J_{n}^{'}\left(nv_{j}\sin\theta\right)J_{n}^{'}\left(nv_{k}\sin\theta\right)\right]\,.
 \end{align}
 We note that when $j=k$, one can resum the result and get the Liénard formula,
 \begin{align}
     P^{\infty, jj}_{LO} =\frac{\left(eQ_{j}\Omega\right)^{2}}{6\pi}\gamma_{j}^{4}v_{j}^{2}\,. 
 \end{align}

\subsection{NLO radiation and energy loss}
While the LO result is the waveform emitted directly from the Schild solution, the NLO contribution, at order $e^3$, is the waveform emitted by the $\mathcal{O}(e^2)$ displacement $z_i^{(1)}(\tau_i)$. We write
$z_j^{\mu,(1)}=\sum_{i=1,2}z_{i\to j}^{\mu}$, where $i$ labels the source and
$j$ the displaced particle. As this displacement involves secular growth, our result for the NLO contribution is only sensible for about a single orbital time period, in which $z^{(1)}_i(\tau_i)$ is still small. To compute the NLO waveform over many time periods, one conventionally works in an \textit{osculating orbit} scheme \cite{Hinderer:2008dm}, in which the energy and angular momentum are given a slow time dependence due to radiation loss. We leave this scheme to future work, and focus on the formally $\mathcal{O}(e^3)$ NLO contribution to the waveform, attaching the radiative propagator as in Eq.~\eqref{eq:Schildrad},
\begin{equation} \label{eq:SchildradNLO}
    {\polvector}_\mu^{\pm} \tilde{A}_{\rm{rad}}^{\mu,(2)} \left(k\right) = -\frac{1}{2}\,\hat{\delta}(k^2)\sgn(k^0)\, i{\polvector}_{\pm}\cdot\left[\mathcal{B}^{0,3}+\mathcal{B}^{1,2}+\mathcal{B}^{3,0}+\mathcal{B}^{2,1}\right]\,.
\end{equation}
Focusing on the $\mathcal{B}^{1,2}$ contributions, we can recast the integral of \eqref{eq:hNLO} into the form 
\begin{equation}\label{eq:B12NLO0}
    \begin{aligned}
i{\polvector}_{\pm}\cdot \mathcal{B}^{1,2}= 
& eQ_2
\int d\tau^b_2 \, e^{ik \cdot \rSchild_2^b}\,\left[(z^b_{1\rightarrow 2}\cdot k)(\uSchild_{2}^{b}\cdot {\polvector}_{\pm})-(k\cdot \uSchild_{2}^{b}) (z^b_{1\rightarrow 2}\cdot{\polvector}_{\pm})\right]\,,
\end{aligned}
\end{equation}
where $z^{b\mu}_{1\rightarrow 2}=z^{\mu}_{1\rightarrow 2}(\tau^b_2)$, and $z^\mu_{1\rightarrow 2}(\tau_2)$ is the displacement of worldline 2 due to a single cut photon emitted from particle 1, satisfying
\begin{eqnarray}\label{eq:Lor12int}
 z_{1\to2}^{\mu}(\tau_2)
 =\int_{-\infty}^{\tau_2}d\tau_2^a\,
   (\tau_2-\tau_2^a)\ddot{z}_{1\to2}^{\mu}(\tau^a_2)\,,
\end{eqnarray}
where
\begin{eqnarray}\label{eq:Lor12}
\ddot{z}_{1\to2}^{\mu}=\frac{eQ_2}{m_2}F_{1\to2}^{\mu\nu}\,
   \mathring u_{2\nu}^a\,.
   \end{eqnarray}
Here the RHS is simply the Lorentz force on particle 2 from half the difference of the retarded and advanced Li\'enard--Wiechert potentials of particle 1.

In \eqref{eq:Lor12} the field tensor is explicitly given by,
\begin{equation}
\begin{aligned}
F^{\mu\nu}_{1\rightarrow 2}
&=\frac{eQ_1}{2}\int \hat{d}^D \ell\,\hat{\delta}\left(\ell^2\right)\sgn(\ell^{0})
   \int d\tau_1 \, e^{i\ell\cdot (\rSchild_1-\rSchild_2)}
   \,\ell^{[\mu}\uSchild_1^{\nu]} \\
&=\frac{eQ_1}{8\pi}
   \left\{\frac{1}{\Delta \rSchild\cdot \uSchild_1}\partial_{\tau_1}
   \left[\frac{\Delta \rSchild^{[\mu} \uSchild^{\nu]}_1}{\Delta \rSchild\cdot \uSchild_1}\right]
   \right\}_{{\rm ret.}+{\rm adv.}}\,,
\end{aligned}
\end{equation}
where we dropped the $a$ superscript for readability. In the bottom line we set $D=4$ and sum over the retarded and advanced values of $\tau_1$, and $\Delta \rSchild=\rSchild_2-\rSchild_1$.  
Evaluating $F^{\mu\nu}_{1\rightarrow 2}$ on the Schild solution (see appendix~\ref{app:Schild}),
\begin{eqnarray}\label{eq:ALD12}
&&\ddot{z}^{\mu}_{1\rightarrow 2}= \frac{eQ_2}{m_2}\mathcal{F}_{1\rightarrow 2}\,\left(\dot{\aSchild}^{\mu}_2-|\aSchild^2_2|\uSchild^{\mu}_2\right)\,,
\end{eqnarray}
where $\aSchild^\mu_j=d\uSchild^\mu_j/d\tau_j$ and,
\begin{eqnarray}\label{eq:f12}
&&\mathcal{F}_{1\rightarrow 2}=\frac{eQ_1v_{1}}{4\pi\gamma^4_2v_2} \frac{\left(1-v_1^2\right)\left(1-v_{2}^2\right) \sin \Phi-\left(\Phi+v_1 v_2 \sin \Phi\right)\left(\cos \Phi+v_1 v_2\right)}{\left(\Phi+v_1 v_2 \sin \Phi\right)^3}\,.
\end{eqnarray}
Note that the relativistic acceleration in Eq.~\eqref{eq:ALD12} has the same form as the Abraham--Lorentz--Dirac force up to overall normalization. In fact, for the diagonal contribution, the self-field subtraction~\eqref{eq:finalH3} gives 
\begin{eqnarray}\label{eq:ALD22}
&&\ddot{z}^{\mu}_{2\rightarrow2}= \frac{e^2Q^2_2}{6\pi m_2}\,\left(\dot{\aSchild}^{\mu}_2-|\aSchild^2_2|\uSchild^{\mu}_2\right)\,,
\end{eqnarray}
where the RHS is the ALD force. Substituting Eq.~\eqref{eq:ALD12} into Eq.~\eqref{eq:Lor12int} we find that the latter formally diverges. This makes sense physically: the Lorentz force has a constant component, and integrating it from $t=-\infty$ leads to infinite displacement. We regulate this divergence by performing a subtraction
\begin{eqnarray}\label{eq:renormz}
 z_{1\to2}^{\mu}(\tau_2)
 \rightarrow\int_{\tau_{*2}}^{\tau_2}d\tau_2^a\,
   (\tau_2-\tau_2^a)\ddot{z}_{1\to2}^{\mu}(\tau^a_2)\,.
\end{eqnarray}
The subtraction fixes the osculation conditions
\begin{eqnarray} \label{eq:SchildOscData}
 z^\mu_{i\to j}(\tau_{*j})=0\,,
 \qquad
 \dot z^\mu_{i\to j}(\tau_{*j})=0\,.
\end{eqnarray}
These conditions are imposed at proper times $\tau_{*j}$ derived from a common coordinate time $t_*$ via
\begin{eqnarray} \label{eq:SchildOscTime}
 \rSchild_i^0(\tau_{*i})=\gamma_i\tau_{*i}=t_*\,,
 \qquad i=1,2\,.
\end{eqnarray}
We then compute the waveform and flux over an interval $\Omega(u-t_*)=\mathcal{O}(1)$ in which the displacement remains perturbative. The resulting displacement is then given by
\begin{eqnarray}\label{eq:z12res}
&&z^{\mu}_{1\rightarrow 2}=\frac{\gamma^2_2eQ_2}{m_2}\mathcal{F}_{1\rightarrow 2}\left[ \uSchild^\mu_2(\tau_2)-T^\mu_2(\tau_2-\tau_{*2})\right]\,,
\end{eqnarray}
where 
\begin{eqnarray} \label{eq:SchildOscTaylor}
T^\mu_j(s)=\uSchild^\mu_j(\tau_{*j})+\aSchild^\mu_j(\tau_{*j})\,s+\frac{v^2_j\gamma^3_j\Omega^2}{2}\hat{t}^\mu\,s^2\,.
\end{eqnarray}
Here and below $\tau_{*j}=t_*/\gamma_j$, so the reference velocity and acceleration in $T_j^\mu$, as well as the explicit acceleration in Eqs.~\eqref{eq:B12NLO0resf} and \eqref{eq:SchildradNLOrx}, are evaluated at the common orbital phase $\Omega t_*$. We note that the part of $z^{b}_{1\rightarrow 2}$ proportional to $\uSchild^{b}_2$ drops out of Eq.~\eqref{eq:B12NLO0}. Consequently, we have
\begin{eqnarray}\label{eq:B12NLO0resf}
i{\polvector}_{\pm}\cdot \mathcal{B}^{1,2}&= 
& \frac{i\gamma^2_2eQ_2}{m_2}\mathcal{F}_{1\rightarrow 2}\left\{-k\cdot T_2\left(-i\gamma^{-1}_2\partial_{k^{0}}-\tau_{*2}\right)\left(i{\polvector}_{\pm}\cdot \mathcal{B}^{0,1}\right)+\left(\aSchild_2(\tau_{*2})\cdot{\polvector}_{\pm}\right)C^{0,1}\right\}\,.\nonumber\\
\end{eqnarray}
Similar expressions hold for the other three contributions to $\mathcal{B}_3$. 
Substituting them into Eq.~\eqref{eq:SchildradNLO} and performing the inverse Fourier transform in the far field limit, we retain twice the outgoing term at nonzero frequency, as in Sec.~\ref{farfieldradLO}. For the zero-frequency term of the potential we use the retarded prescription. We obtain,
\begin{eqnarray} \label{eq:SchildradNLOrx}
    &&{\polvector}_\mu^{\pm} A_{\rm{out}}^{\mu,(2)} \left(x\right) =
-\sum_{i, j} \frac{\gamma_je^2Q^2_j}{4\pi m_j r}\mathcal{F}_{i\rightarrow j}\sum_{n=-\infty}^{\infty}\,(-i)^ne^{in\bar{\varphi}_j}\left\{\left(\aSchild_j(\tau_{*j})\cdot{\polvector}_{\pm}\right) \, J_{n}\left(k^{0} R_j \sin\theta\right) \,e^{-i k^{0} u}\right.\nonumber\\[5pt]
&&\left.-\frac{i\gamma_js_jv_j}{\sqrt{2}}e^{\pm i \bar{\varphi}^j}q\cdot T_j\left(i\gamma^{-1}_j\partial_{k^{0}}-\tau_{*j}\right)\left[  \, J_{n\pm 1}\left(k^{0} R_j \sin\theta\right)\,k^{0} \,e^{-i k^{0} u}\right]\right\}_{k^{0}=n\Omega},
\end{eqnarray}
where $q^\mu=(1,\hat{\boldsymbol x})$. The NLO potential contains terms
linear and quadratic in $u-t_*$. By \eqref{eq:SchildOscData}, the contribution of
particle $j$ to the NLO potential vanishes at the retarded time
\begin{eqnarray*}
 u=q_\mu\rSchild_j^\mu(\tau_{*j})\,,
 \qquad \tau_{*j}=t_*/\gamma_j\,.
\end{eqnarray*}
The two particles generally have different arrival times for a given
observation direction. The resulting far-field electric field is
\begin{eqnarray}\label{eq:E2}
\boldsymbol{E}_{\mathrm{out}}^{\perp,(2)}&= &\partial_u\left[\left({\polvector}^{-}\cdot A_{\rm{out}}^{(2)}\right) \bar{\vec{\polvector}}^++\left({\polvector}^{+}\cdot A_{\rm{out}}^{(2)}\right) \bar{\vec{\polvector}}^-\right]\,.
\end{eqnarray}
The NLO flux comes from the mixed term between $\boldsymbol{E}_{\mathrm{out}}^{\perp,(1)}$ and $\boldsymbol{E}_{\mathrm{out}}^{\perp,(2)}$, namely
\begin{eqnarray} \label{eq:ElossNLO}
    P^{\infty}_{\rm NLO} = 2\int d\Omega\,r^{2}\,\vec{E}_{\rm{out}}^{\perp,(1)}\cdot \vec{E}_{\rm{out}}^{\perp,(2)}\,.
\end{eqnarray}
We omit the explicit expression for brevity, but it can be obtained directly by substituting Eqs.~\eqref{eq:E1} and \eqref{eq:E2} into Eq.~\eqref{eq:ElossNLO}.
\subsection{Non-relativistic expansion}
To cross-check our relativistic results \eqref{eq:LOflux} and \eqref{eq:ElossNLO}, we expand them to low orders in $v/c$, before comparing them in the next section to standard non-relativistic EFT. 
To streamline the presentation, we define the combinations
\begin{eqnarray}
&&D_a  =\sum_j Q_j s_j v_j^a ~~~~~\quad a\in\{1,3\},~~~~~~~  S_2  =\sum_j Q_j v_j^2, \\
&&H_a  =\sum_j Q_j^2 v_j^a / m_j \quad a\in\{0,2\}, ~~~~~~~ J_1 =\sum_j Q_j^2 s_j v_j / m_j \,.
\end{eqnarray}

Defining $\psi\equiv\varphi-\Omega u$, the velocity-expanded \eqref{eq:E1} is
\begin{eqnarray}\label{eq:E1v}
    &&\vec{E}^{\perp,(1)}_{\rm{out}} =\frac{e \Omega}{4 \pi \sqrt{2} r}\,\left\{e^{-i \Omega u}\,\mathcal{W}_{(1)}\,\bar{\vec{\polvector}}^++{\rm cc.}\right\}+\mathcal{O}\left(v^4\right)\,,
\end{eqnarray}
where
\begin{eqnarray}
\mathcal{W}_{(1)}=D_1-2 i S_2 \sin \theta e^{i \psi}+\frac{D_3 \sin ^2 \theta}{8}\left(e^{-2 i \psi}-2-27 e^{2 i \psi}\right) \,.
\end{eqnarray}
Substituting Eq.~\eqref{eq:E1v} into Eq.~\eqref{eq:energyloss}, we get the velocity expanded LO flux,
\begin{eqnarray}\label{eq:dEdtLOv}
P^\infty_{\rm LO}=\frac{e^2 \Omega^2}{4 \pi}\left(\frac{2}{3} D_1^2-\frac{4}{15} D_1 D_3+\frac{8}{5} S_2^2\right)+\mathcal{O}\left(v^6\right) \,.
\end{eqnarray}
The leading-in-$v$ contribution to the NLO field is
\begin{eqnarray}\label{eq:E2v}
    &&\mathbf{E}_{\mathrm{out}}^{\perp,(2)}=\frac{e^3 \Omega^2}{24 \sqrt{2} \pi^2 r}\left\{e^{-i \Omega u} \,\mathcal{W}_{(2)} \,\bar{\vec{\polvector}}^++\text {c.c. }\right\}+\mathcal{O}\left(v^4\right)\,,\nonumber\\
\end{eqnarray}
where 
\begin{eqnarray}\label{eq:E2vW}
\mathcal{W}_{(2)}  =D_1\,\left(-i H_0+J_1 \beta_1+H_2 \beta_2+\frac{7i}{10} H_2\right)+\frac{i}{5} \left(D_3 H_0-12 S_2 J_1\right),
\end{eqnarray}
and
\begin{eqnarray}\label{eq:B12}
 \beta_1
 &=& \sin\theta\left[
 (2e^{iw}-4)e^{i\psi}
 +e^{-i\psi}\cos w
 -iw\cos(\psi+w)
 \right]\,,
 \\[5pt]
 \beta_2
 &=& 2w+i\left(1-\frac{w^2}{2}\right)
 +\frac{i\sin^2\theta}{8}\Bigg\{
 e^{2i\psi}\left[81-16(3-iw)e^{iw}\right]
 \nonumber\\
 &&\qquad
 +e^{-2i\psi}\left(1+8e^{iw}\right)
 +2-8(1-2iw)e^{-iw}
 \Bigg\}\,.
\end{eqnarray}
In the above expression we defined the shorthand $w=\Omega(u-t_*)$.
Substituting Eqs.~\eqref{eq:E2v} and \eqref{eq:E1v} into Eq.~\eqref{eq:ElossNLO}, we get the leading-in-$v$ NLO flux,
\begin{eqnarray}\label{eq:dEdtNLOv}
P^\infty_{\rm NLO}=\frac{e^4 \Omega^3}{18 \pi^2}\left[D_1^2 H_2 F(w)+D_1 S_2 J_1 G(w)\right]+\mathcal{O}\left(v^6\right)\,,
\end{eqnarray}
with
\begin{eqnarray}\label{eq:fluxNLOFG}
 F(w) &=& 2w-\frac{6w\cos w+\sin w}{5}\,,
 \nonumber\\[3pt]
 G(w) &=& \frac35\left(w\cos w-4\sin w\right)\,.
\end{eqnarray}

\subsection{Benchmark vs. non-relativistic moments}
\label{sec:benchmark}
We can now benchmark our $v$-expanded results \eqref{eq:dEdtLOv} and \eqref{eq:dEdtNLOv} against standard electromagnetic results. The latter are based on the electromagnetic multipole emission formula (in its instantaneous version)  \cite{Ross:2012fc},
\begin{eqnarray}\label{eq:momloss}
&&P^\infty =\frac{1}{4\pi}\left\{\frac{2}{3}\tddot{\mathcal{I}}_a\tddot{\mathcal{I}}_a+\frac{1}{20}\tdddot{\mathcal{I}}_{ab}\tdddot{\mathcal{I}}_{ab}+\mathcal{O}(v^6)\right\}\,,
\end{eqnarray}
where $a,b$ are Cartesian indices, primes denote coordinate time derivatives, and $\mathcal{I}_a$ and $\mathcal{I}_{ab}$ are the radiative electric dipole and quadrupole moments, respectively,
\begin{eqnarray}\label{eq:moments}
&&\mathcal{I}_a=\int d^3 \vec{x} \left\{j^0 \vec{x}_a +\frac{1}{30} \frac{d^2}{d t^2} \left(j^0 r^2 \vec{x}_a\right)+\frac{1}{6} \frac{d}{d t} \left[r^2 \vec{j}_a-(\vec{j} \cdot \vec{x}) \vec{x}_a\right]+\ldots\right\} \,,\nonumber\\[5pt]
&&
\mathcal{I}_{a b}=  \int d^3 \vec{x} \left\{j^0 \vec{x}_{\langle a} \vec{x}_{b\rangle}+\ldots\right\} \,,
\end{eqnarray}
where $A_{\langle a} B_{b\rangle} \equiv \frac{1}{2}\left(A_a B_b+A_b B_a\right)-\frac{1}{3} \delta_{ab}(A \cdot B)$. The currents are given by
\begin{eqnarray}\label{eq:Schildj}
j^0(t, \mathbf{x})= e \sum_{i=1,2}Q_i \delta^{(3)}\left[\mathbf{x}-\vec{x}_i(t)\right]~~,~~
\mathbf{j}(t, \mathbf{x})= e \sum_{i=1,2}Q_i \vec{v}_i(t) \delta^{(3)}\,\left[\vec{x}- \vec{x}_i(t)\right]\, .
\end{eqnarray}
The leading magnetic dipole is constant on the circular reference orbit, so its contribution to the LO power and to the LO--NLO interference vanishes. To $\mathcal{O}(e^2)$ (NLO), the trajectories are given by
\begin{eqnarray}\label{eq:traj}
\vec{x}_i=s_iR_i\vhrSchild+\sum_{j}\delta \vec{x}_{j\rightarrow i}~~,~~\vec{v}_i=s_iv_i\vhvSchild+\sum_{j} \frac{d}{dt}\delta \vec{x}_{j\rightarrow i}\,,
\end{eqnarray}
where $\vhrSchild(t)=(\cos\Omega t,\sin\Omega t,0)$ and  $\vhvSchild(t)=(-\sin\Omega t,\cos\Omega t,0)$ are the unit vectors of the Schild solution. The equal-time displacement $\delta\vec{x}_{j\rightarrow i}$ due to photon exchange will be computed below. 
To get the NLO moments we first expand the currents \eqref{eq:Schildj} to linear order in $\delta\vec{x}_{j\rightarrow i}(t)$, and get
\begin{eqnarray}
j^0(t,\vec{x})
=j^0_{\mathrm{LO}}(t,\vec{x})+j^0_{\mathrm{NLO}}(t,\vec{x})~~,~~\vec{j}(t,\vec{x})
=\vec{j}_{\mathrm{LO}}(t,\vec{x})+\vec{j}_{\mathrm{NLO}}(t,\vec{x})\,,
\end{eqnarray}
where
\begin{eqnarray}
j^0_{\mathrm{LO}}(t,\vec{x}) 
&=& e\sum_i Q_i
 \delta^{(3)}\!\left[\vec{x}-s_iR_i\vhrSchild(t)\right] \,, \nonumber\\[5pt]
\vec{j}_{\mathrm{LO}}(t,\vec{x})&=& e\sum_i Q_i
 s_iv_i\vhvSchild(t)\delta^{(3)}\!\left[\vec{x}-s_iR_i\vhrSchild(t)\right]\,,
\end{eqnarray}
and so
\begin{eqnarray}\label{eq:ILO}
\mathcal{I}^{\mathrm{LO}}_a&=&\frac{e}{\Omega}\left(D_1-\frac{D_3}{5}\right)\vhrSchild_a(t)+\mathcal{O}\left(v^5\right)\,,\nonumber\\[5pt]
\mathcal{I}_{ab}^{\mathrm{LO}}(t)
&=& \frac{e S_2}{\Omega^2}\,
  \vhrSchild_{\langle a}(t)\,\vhrSchild_{b\rangle}(t)
  + \mathcal{O}\!\left(v^4\right).
\end{eqnarray}
Taking time derivatives, we get
\begin{eqnarray}\label{eq:ddILO}
\tddot{\mathcal{I}}^{\mathrm{LO}}_a&=&-e\Omega\left(D_1-\frac{D_3}{5}\right)\vhrSchild_a(t) + \mathcal{O}\left(v^5\right)\,,\nonumber\\[5pt]
\tdddot{\mathcal{I}}_{ab}^{\mathrm{LO}}
&=& -8e\Omega S_2\,
   \vhrSchild_{\langle a}(t)\,\vhvSchild_{b\rangle}(t)
   + \mathcal{O}\!\left(v^4\right).
\end{eqnarray}
Upon substituting in \eqref{eq:momloss}, these reproduce the leading-in-$v$ LO flux \eqref{eq:dEdtLOv} exactly.
The NLO contributions to the currents are
\begin{equation}
\begin{aligned}
\label{eq:curNLO}
j^0_{\mathrm{NLO}}(t,\vec{x})
=& -e\sum_i Q_i
 \left[\sum_j\delta\vec{x}_{j\rightarrow i}(t)\cdot\vec{\nabla}_{x}\right]
 \delta^{(3)}\!\left[\vec{x}-s_i R_i\vhrSchild(t)\right]\,,
\\[5pt]
\vec{j}_{\mathrm{NLO}}(t,\vec{x})
=& e\sum_i Q_i\Bigg\{
 -s_iv_i\vhvSchild(t)
 \sum_j\delta\vec{x}_{j\rightarrow i}(t)\cdot\vec{\nabla}_{x}
 +\sum_j\frac{d}{dt}\delta \vec{x}_{j\rightarrow i}(t)
 \Bigg\}
 \delta^{(3)}\!\left[\vec{x}-s_iR_i\vhrSchild(t)\right]\,.
\end{aligned}
\end{equation}
The equal-time displacement is defined as
\begin{eqnarray}
\delta x^\mu_{j\rightarrow i}(t)
\equiv \left(0,\delta\vec{x}_{j\rightarrow i}(t)\right)
=\rSchild^\mu_i(\tau+\delta\tau)+z^\mu_{j\rightarrow i}(\tau+\delta\tau)
 -\rSchild^\mu_i(\tau)\,,
\end{eqnarray}
where $t=\rSchild_i^0(\tau)$ and $\delta\tau$ is chosen so that
$\delta x^0_{j\rightarrow i}=0$. To first order in $z$, this gives
\begin{eqnarray}\label{eq:Dxij}
\delta\vec{x}_{j\rightarrow i}(t)
=\boldsymbol{z}_{j\rightarrow i}(\tau)
 -\frac{d\vrSchild_i(t)}{dt}\,z^0_{j\rightarrow i}(\tau)\,.
\end{eqnarray}

By~\eqref{eq:SchildOscTime}, the two reference proper times correspond to the same
center-of-mass coordinate time,
\begin{eqnarray}
 \gamma_i\tau_{*i}=t_*\,,\qquad i=1,2\,.
\end{eqnarray}
The osculation conditions~\eqref{eq:SchildOscData} therefore give
\begin{eqnarray}
 \delta\vec x_{j\rightarrow i}(t_*)=0\,,\qquad
 \left.\frac{d}{dt}\delta\vec x_{j\rightarrow i}(t)\right|_{t=t_*}=0\,.
\end{eqnarray}
Using Eqs.~\eqref{eq:f12}--\eqref{eq:SchildOscTaylor} and \eqref{eq:Dxij}, the displacement through cubic order in the speeds is
\begin{eqnarray}
 \delta \vec x_{j\rightarrow i}(t)
 &=&\frac{e^2Q_iQ_js_jv_j}{6\pi m_i}\Bigg\{
 \left(1-\frac7{10}v_i^2-\frac15v_j^2
            +\frac{12}{5}s_is_jv_iv_j\right)
 \nonumber\\
 &&\times\left[\vhvSchild(t)-\vhvSchild(t_*)
               +\Omega(t-t_*)\vhrSchild(t_*)\right]
 +\frac12v_i^2\Omega^2(t-t_*)^2\vhvSchild(t)
 \Bigg\} + \mathcal{O}(v^5).
\end{eqnarray}
Substituting the displacement into Eq.~\eqref{eq:curNLO} and then the currents into Eq.~\eqref{eq:moments}, we obtain the radiative moments at $u$, with $w=\Omega(u-t_*)$,

\begin{eqnarray}
 \mathcal I_a^{\mathrm{NLO}}(u)
 &=&\frac{e^3}{6\pi}\Bigg\{
 \left(D_1H_0-\frac7{10}D_1H_2-\frac15D_3H_0+\frac{12}{5}S_2J_1\right)
 \nonumber\\
 &&\qquad\times
 \big[\vhvSchild_a(u)-\vhvSchild_a(t_*)+w\vhrSchild_a(t_*)\big]
 \nonumber\\
 &&\quad+\frac{D_1H_2}{10}\big[
 (5w^2-2-3w\sin w)\vhvSchild_a(u)
 +(5\sin w-3w\cos w)\vhrSchild_a(u)\big]\Bigg\}
 +\mathcal{O}(v^5)\,,
 \nonumber\\[4pt]
 \mathcal I_{ab}^{\mathrm{NLO}}(u)
 &=&\frac{e^3D_1J_1}{3\pi\Omega}\Big[
 \vhrSchild_{\langle a}(u)\vhvSchild_{b\rangle}(u)
 -\vhrSchild_{\langle a}(u)\vhvSchild_{b\rangle}(t_*)
 + w\vhrSchild_{\langle a}(u)\vhrSchild_{b\rangle}(t_*)\Big]
 +\mathcal{O}(v^4)\,.
 \label{eq:INLO}
\end{eqnarray}
Differentiating Eq.~\eqref{eq:INLO} with respect to $u$ at fixed $t_*$ gives
\begin{eqnarray}
 \tddot{\mathcal I}_a^{\mathrm{NLO}}(u)
 &=& \frac{e^3\Omega^2}{12\pi}\Bigg\{
 \Big[-2D_1H_0+\frac25(D_3H_0-12S_2J_1)
 \nonumber\\
 &&\qquad +\frac{D_1H_2}{5}
 \big(19-5w^2+12w\sin w-2\cos w\big)\Big]\vhvSchild_a(u)
 \nonumber\\
 &&\qquad -2D_1H_2\left[2w-\frac{6w\cos w+\sin w}{5}\right]
 \vhrSchild_a(u)\Bigg\}+\mathcal{O}(v^5)\,,
 \nonumber\\[4pt]
 \tdddot{\mathcal I}_{ab}^{\mathrm{NLO}}(u)
 &=& \frac{e^3\Omega^2D_1J_1}{3\pi}\Big[
 4\vhrSchild_{\langle a}(u)\vhrSchild_{b\rangle}(u)
 -4\vhvSchild_{\langle a}(u)\vhvSchild_{b\rangle}(u)
 \nonumber\\
 &&\qquad
 +\vhvSchild_{\langle a}(u)\vhvSchild_{b\rangle}(t_*)
 -3\vhrSchild_{\langle a}(u)\vhrSchild_{b\rangle}(t_*)
 -w\vhvSchild_{\langle a}(u)\vhrSchild_{b\rangle}(t_*)
 \Big]+\mathcal{O}(v^4)\,.
 \label{eq:ddINLO}
\end{eqnarray}
In Eq.~\eqref{eq:momloss} the interference with the LO moment derivatives \eqref{eq:ddILO}, evaluated at $u$, gives
\begin{eqnarray}
 \frac1{3\pi}{\tddot{\mathcal I}}_a^{\mathrm{LO}}{\tddot{\mathcal I}}_a^{\mathrm{NLO}}
 &=&\frac{e^4\Omega^3}{18\pi^2}D_1^2H_2\,F(w)+\mathcal{O}(v^6)\,,\nonumber\\
 \frac1{40\pi}\tdddot{\mathcal I}_{ab}^{\mathrm{LO}}\tdddot{\mathcal I}_{ab}^{\mathrm{NLO}}
 &=&\frac{e^4\Omega^3}{18\pi^2}D_1S_2J_1\,G(w)+\mathcal{O}(v^6)\,.
\end{eqnarray}
Their sum reproduces the instantaneous NLO flux \eqref{eq:dEdtNLOv} through $\mathcal{O}(v^4)$.

\section{Discussion}
\label{sec:concl}

Methods from quantum scattering amplitudes have had substantial impact on
the computation of nonradiative and radiative observables for classical
gravity.  The direct application of these methods to date has largely
been in computing observables in the unbound (scattering) domain.  Observations, in contrast, are exclusively of binary systems in 
circular motion.  In this paper, we have described an approach and
an algorithm for bridging this gap.

In more detail, we presented the calculation of a waveform radiated from a bound state in electrodynamics.  
The bound state solution we use is the circular Schild solution to Wheeler--Feynman electrodynamics: a time-symmetric formulation of electrodynamics.
Consequently the Schild solution is a nonradiating starting point for a perturbative expansion.  
We presented the computation of the NLO corrections to the waveform around
this solution.

In our approach, the key to mapping unbound (scattering) quantities to bound-state ones is to work with \textit{integrands\/}.  
We showed how to map a slightly modified unbound integrand --- given by a scattering amplitude --- to the corresponding integrand for a bound-state.  
The mapping is independent of the specific solution we are using.  
We have checked the correctness and consistency by comparing with a worldline field theory calculation in the background of the Schild solution, to next-to-leading order and, for some contributions, to one higher order.

There is a time-symmetric theory of gravity~\cite{Hoyle:1964} analogous to the Wheeler--Feynman formulation of electrodynamics.
It has a solution (due to Friedman and Uryu~\cite{Friedman:2005rx}) which is similar to the Schild solution. 
Naturally we look forward to applying our approach to gravity based on this formulation.
It would also be interesting to generalise the Schild solution away from circular motion.

As mentioned briefly in Section~\ref{sec:pertSetup}, the Wheeler--Feynman boundary conditions can be understood as arising from an incoming radiation field which is precisely tuned to balance the loss of energy in ordinary electrodynamics.
It would be interesting to understand this setup in the full quantum theory with a possible role for an incoming coherent state.
In linearised gravity a similar approach to the Wheeler--Feynman boundary conditions should be available.
It would be interesting to study an extension to full gravity.
Finally, the links between the ideas presented here and the all-orders structure recently exposed in Ref.~\cite{Bautista:2026ozj} merit further study.

\acknowledgments
We thank G.~Brown, S.~De Angelis, and M.~Zeng for discussions.
M.K. acknowledges partial support from the ERC Starting Grant ``Light-Dark'' (Grant No. 101040019), and the Azrieli Fellowship. O.T. is supported in part by the ISF grant No. 3533/24, the BSF grant No. 2024169, and by grant No. 2025785 from the US-Israel BSF/NSF.
D.A.K.'s research was supported by the European Research Council (ERC) under the European Union's research and innovation program grant agreement ERC--AdG--885414 (`Ampl2Einstein').
Funded by the European Union (ERC, Classical$+$, 101200505). 
For the purpose of open access, the authors grant a Creative Commons Attribution (CC BY) license to any Author Accepted Manuscript version arising from this submission.
This project has received funding from the European Research Council (ERC) under the European Union’s Horizon Europe research and innovation programme (grant agreement No. 101040019). Views and opinions expressed are however those of the author(s) only and do not necessarily reflect those of the European Union. The European Union cannot be held responsible for them.

\paragraph{Use of AI} 
The form of Eq.~\eqref{eq:SchildSubtractedAction} was suggested in interactions with Gemini 3.1 Pro.
We further used GPT 6.2 Astra to audit the paper for linguistic errors, mathematical typos, and also the non-relativistic benchmarking of our results.

\appendix

\section{The Schild solution}\label{app:Schild}

\begin{figure}[h]
  \centering
  \begin{tikzpicture}[line cap=round, line join=round]
    % Schematic: radius ratio and angle are illustrative only.
    \def\rinner{1.5}   % orbit of m_2 (radius v_2/Omega)
    \def\router{2.5}   % orbit of m_1 (radius v_1/Omega)
    \def\tilt{15}    % instantaneous position of m_2
    \def\phival{45}  % retardation angle Phi (degrees)

    \coordinate (O)    at (0,0);
    \coordinate (M1)   at ({\tilt+180}:\router);
    \coordinate (M2)   at (\tilt:\rinner);
    \coordinate (Mret) at ({\tilt-\phival}:\rinner); % behind (CCW motion)
    \coordinate (Madv) at ({\tilt+\phival}:\rinner); % ahead

    \draw[blue, line width=1pt] (O) circle (\router);
    \draw[red,  line width=1pt] (O) circle (\rinner);
    \draw[->, thick] ({\tilt+100}:{\router+0.3})
      arc[start angle={\tilt+100}, end angle={\tilt+130}, radius={\router+0.3}];
    \node at ({\tilt+115}:{\router+0.75}) {$\Omega$};

    \draw[densely dotted, line width=0.8pt] (O) -- (M2);
    \draw[line width=0.9pt] (M1) -- (O);
    \draw[line width=0.9pt] (Mret) -- (O) -- (Madv);
    \draw[dashed, gray] (Mret) -- (M1) -- (Madv);   % null rays

    \draw[thin] (\tilt:0.85)
      arc[start angle=\tilt, end angle={\tilt+\phival}, radius=0.85];
    \node at ({\tilt+\phival/2}:1.10) {$\Phi$};
    \draw[thin] (\tilt:1.25)
      arc[start angle=\tilt, end angle={\tilt-\phival}, radius=1.25];
    \node at ({\tilt-\phival/2}:1) {$\Phi$};

    \fill[red] (Madv) circle (1.7pt);
    \node[red] at ({\tilt+\phival+9}:{(\rinner+\router)/2.2}) {$m_{2,\mathrm{adv}}$};
    \fill[red] (Mret) circle (1.7pt);
    \node[red] at ({\tilt-\phival-9}:{(\rinner+\router)/2}) {$m_{2,\mathrm{ret}}$};
    \draw[red, fill=white] (M2) circle (1.7pt);
    \fill[blue] (M1) circle (1.7pt) node[below left, inner sep=3pt] {$m_1$};
  \end{tikzpicture}
  \caption{Snapshot of the Schild orbits at time $t$ in the CM frame; both particles rotate anticlockwise with angular velocity $\Omega$. 
    Particle 1 (blue, radius $v_1 / \Omega$) interacts with particle 2 (red, radius $v_2 / \Omega$) not at its instantaneous, diametrically opposite position (open circle), but at its retarded and advanced positions, displaced by an angle $\Phi$ on either side. Dashed lines are the connecting null rays, of length $\Phi/\Omega$.}
  \label{fig:schild}
\end{figure}
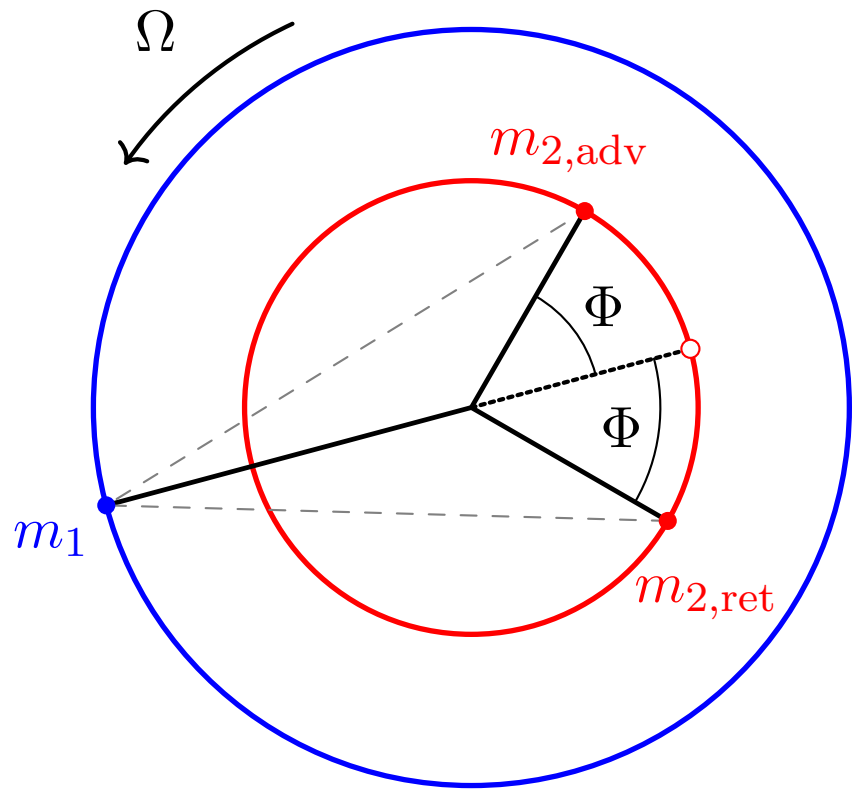

Schild's circular orbits~\cite{Schild:1963} solve the time-symmetric Wheeler--Feynman equations of motion, in which the force on each particle depends on the other particle's worldline both in its past and in its future. 
Rather than integrating the equations of motion from initial data, we therefore make an ansatz for the complete worldlines and treat the equations of motion as consistency conditions on the ansatz parameters.
Working in the center-of-momentum (CM) frame, we require both particles to move on concentric circles in the $xy$-plane with common
angular velocity $\Omega$, on diametrically opposite sides of the origin. 
We write the orbits as:
\begin{equation}
\begin{aligned}
\label{eq:schildStarterapp}
  \rSchild^\mu_1(\tau)&=\Big(\gamma_1 \tau_1,\,
  \Omega^{-1}v_1\cos(\Omega \, \gamma_1\tau_1),\,
  \Omega^{-1}v_1\sin(\Omega \, \gamma_1\tau_1),\,0\Big) \,, \\
  \rSchild^\mu_2(\tau)&=\Big(\gamma_2 \tau_2,\,
  -\Omega^{-1}v_2\cos(\Omega \, \gamma_2\tau_2),\,
  -\Omega^{-1}v_2\sin(\Omega \, \gamma_2\tau_2),\,0\Big)\,.
\end{aligned}
\end{equation}
where, as usual, the gamma factors are
\begin{equation}
\gamma_i\coloneqq\big(1-v_i^2\big)^{-1/2} \,.
\end{equation}
The orbital radii are $R_i = v_i / \Omega$ for each particle. 
As appropriate for circular motion, $R_i$ and $\Omega$ are constants.
The geometry of the orbits is sketched in Fig.~\ref{fig:schild}.

In accordance with the time-symmetric boundary conditions, the force acting on particle 1 is due to the average field of particle 2 at \emph{both} its retarded and advanced positions (Fig.~\ref{fig:schild}).
With $t_1=\gamma_1\tau_1$, retarded $t_2^{\rm ret}$ and advanced $t_2^{\rm adv}$ times  are fixed by requiring 
\begin{equation}
\label{eq:schildTimeDelay}
\big(\rSchild_2(\tau_2)-\rSchild_1(\tau_1)\big)^2=0 \,.
\end{equation}
This equation has two solutions for $\tau_2$, corresponding to times $t_2 = \gamma_2 \tau_2$ which are either retarded or advanced according to $t_2^{\rm ret}<t_1<t_2^{\rm adv}$.
It is helpful to define the retardation parameter
\begin{equation}
  \Phi\equiv\Omega\big(t_1-t_2^{\rm ret}\big)=\Omega\big(t_2^{\rm adv}-t_1\big)>0\,,
\end{equation}
i.e.\ the angle swept out by particle 2 during the light-travel time. 
In view of the symmetry of the solution, the same $\Phi$ describes both retarded and advanced points. 
The geometry of the time delay (Fig.~\ref{fig:schild}), or equivalently expanding~\eqref{eq:schildTimeDelay}, gives
\begin{equation}\label{eq:retcond}
  \Phi^2=v^2_1+v^2_2+2v_1v_2\cos\Phi\,.
\end{equation}

By the symmetry of the configuration only the radial component of the equations of motion is non-trivial: the tangential forces from the retarded and advanced images cancel in the time-symmetric combination, the $z$-component vanishes for planar motion, and the time component is proportional to $\mathbf F_i\cdot\vec{v}_i=0$, where $\mathbf{F}_i$ is the force on particle $i$ and $\vec{v}_i$ is its spatial (coordinate) velocity. 
Substituting \eqref{eq:schildStarterapp} into the (conservative) Lorentz force of Eq.~\eqref{eq:conservativeEOM} 
then leads to the relations:
\begin{equation}\label{eq:SchildEOMapp}
\begin{aligned}
  \frac{m_1\gamma_1 v_1}{\Omega}
  &=\frac{e^2| Q_1 Q_2|}{4\pi\,(\Phi+v_1 v_2\sin\Phi)^3}
  \Big[(v_1+v_2\cos\Phi)\big(1-v^2_1\big)\big(1-v^2_2\big)\\
  &\hspace{3.2cm}+(v_1\Phi+v_2\sin\Phi)(\Phi+v_1 v_2\sin\Phi)\Big]\,,\\
  \frac{m_2\gamma_2 v_2}{\Omega}
  &=\frac{e^2| Q_1 Q_2|}{4\pi\,(\Phi+v_1 v_2\sin\Phi)^3}
  \Big[(v_2+v_1\cos\Phi)\big(1-v^2_1\big)\big(1-v^2_2\big)\\
  &\hspace{3.2cm}+(v_2\Phi+v_1\sin\Phi)(\Phi+v_1 v_2\sin\Phi)\Big]\,,
\end{aligned}
\end{equation}
where, for generality, we have written the charges of the two particles as $Q_i$ for $i=1,2$ (of course, the charges must have opposite signs to bind the particles together).
These equations can be viewed as relativistic, conservative versions of Kepler's third law (or rather of the electromagnetic analogue).
Together with Eq.~\eqref{eq:retcond} they determine $v_1$, $v_2$ and $\Phi$ for given $m_1$, $m_2$ and $\Omega$.

\section{Locating Required Subtractions}
\label{app:MappingProcedure}

\begin{figure}[b]
\centering
\begin{tikzpicture}[scale=1.0, baseline={([yshift=-\the\dimexpr\fontdimen22\textfont2\relax]current bounding box.center)},
        ] 
\begin{feynman}
  \vertex (v) {};
  \vertex [left=2 of v] (il);
  \vertex [left=0.75 of v] (vil);
  \vertex [right=0.75 of v] (vol);
  \vertex [right=2 of v] (ol);
  \vertex [above=1.5 of v] (vu);
  \vertex [above=1.5 of vil] (viu);
  \vertex [above=1.5 of vol] (vou);
  \vertex [above=1.5 of il] (iu);
  \vertex [above=1.5 of ol] (ou);
  \vertex [below left=1 and 0.2 of ol] (ogamma);
  \draw [thick] (iu) -- (ou);
  \draw [thick] (il) -- (ol);
  \diagram* {
    (viu) -- [photon, thick] (vil);
    (vou) -- [photon, thick] (vol);
    (v) -- [photon, thick] (ogamma);
  };
  \filldraw [fill=allOrderBlue] (v) ellipse (25pt and 10pt);
\end{feynman}
\end{tikzpicture}
\caption{A five-point loop amplitude contribution to the waveform at $\Ord(Q_1^2 Q_2^3)$.}
\label{fig:NLOContribution}
\end{figure}
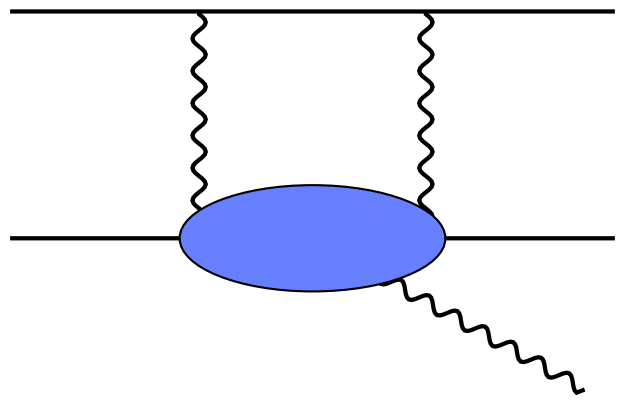

We give an example of how to identify the terms potentially
requiring subtractions
on photon propagators in a $\Ord(Q_1^2 Q_2^3)$ contribution to
the waveform, shown in Fig.~\ref{fig:NLOContribution}.  We focus
on the lower amplitude obtained by cutting all photon propagators;
the required loop amplitude can be recovered by sewing this
amplitude together with the upper amplitude, using the appropriate
unsubtracted or subtracted photon propagators as identified here.
Terms lacking a maximal cut will give rise only to contact terms,
and will not contribute to the long-distance behavior in coordinate
space in which we are ultimately interested.  For simplicity
of exposition, we drop the decorations here; but the procedure
should be carried out on the decorated amplitude.

\newcommand{\pe}[1]{p\!\cdot\!\polvector_{#1}}
\newcommand{\llp}[2]{\ell_{#1}\!\cdot\!\ell_{#2}}
\newcommand{\lep}[2]{\ell_{#1}\!\cdot\!\polvector_{#2}}
\newcommand{\ee}[2]{\polvector_{#1}\!\cdot\!\polvector_{#2}}
\newcommand{\adenom}[1]{(p\cdot\ell_{#1})}

The classical limit of the three-photon two-scalar amplitude (with incoming photon momenta $\ell_1$ and $\ell_2$ and outgoing momentum $\ell_3$, and with an incoming massive particle $p$) is given by,
\begin{equation}
\begin{aligned}
\label{eq:fullAmp5}
A_5=&\left[\ee{1}{2}-\lep{2}{1}\frac{\pe{2}}{\adenom2}
-\lep{1}{2}\frac{\pe{1}}{\adenom1}
+\llp{1}{2}\frac{\pe{1}}{\adenom1}\frac{\pe{2}}{\adenom2}\right]\\
&\quad\times\left[-\frac{(l_1+l_2)\cdot\polvector_3^*}{\adenom3}
+(l_1+l_2)\cdot l_3\,\frac{p\cdot\polvector_{3}^*}{\adenom3^2}\right]\\[8pt]
&+\left[\polvector_{3}^*\cdot\polvector_{1}-\lep{1}{3}^*\frac{\pe{1}}{\adenom1}
-\lep{3}{1}\frac{\pe{3}^*}{\adenom3}
+\llp{3}{1}\frac{\pe{3}^*}{\adenom3}\frac{\pe{1}}{\adenom1}\right]\\
&\quad\times\left[\frac{(-l_3+l_1)\cdot\polvector_2}{\adenom2}
-(-l_3+l_1)\cdot l_2\,\frac{\pe{2}}{\adenom2^2}\right]\\[8pt]
&+\left[\ee{2}{3}^*-\lep{3}{2}\frac{\pe{3}^*}{\adenom3}
-\lep{2}{3}^*\frac{\pe{2}}{\adenom2}
+\llp{2}{3}\frac{\pe{2}}{\adenom2}\frac{\pe{3}^*}{\adenom3}\right]\\
&\quad\times\left[\frac{(l_2-l_3)\cdot\polvector_1}{\adenom1}
-(l_2-l_3)\cdot l_1\,\frac{\pe{1}}{\adenom1^2}\right]\,.
\end{aligned}
\end{equation}

In order to identify the `double pole' terms, we first extract
the leading pole terms, those with an overall scaling of
$(\adenom{i})^{-4}$.  In this example, we obtain,
\begin{equation}
\begin{aligned}
\pe1\,\pe2\,\pe3^*\Bigg[&\frac{\llp{1}{2}\,\llp{1}{3}}{\adenom1\adenom2\adenom3^2}
+\frac{\llp{1}{2}\,\llp{2}{3}}{\adenom1\adenom2\adenom3^2}\\
&+\frac{\llp{1}{3}\,\llp{2}{3}}{\adenom1\adenom3\adenom2^2}
-\frac{\llp{1}{3}\,\llp{1}{2}}{\adenom1\adenom3\adenom2^2}\\
&-\frac{\llp{2}{3}\,\llp{1}{2}}{\adenom1^2\adenom2\adenom3}
+\frac{\llp{2}{3}\,\llp{1}{3}}{\adenom1^2\adenom2\adenom3}\Bigg].
\end{aligned} \label{eq:pole-seed}
\end{equation}

In the worldline theory, each denominator factor can appear with
at most a quadratic power.  This means that the amplitude can be
cast in the same form.  We write down an ansatz with each possible
product of $n-1$ 
squared eikonalized matter denominators for an $n$-photon amplitude.
(There are $n$ different possibilities.)
In the specific example at hand, each denominator is a
product of two squared eikonalized matter denominators,
\begin{equation}
    \adenom1^2\,\adenom2^2\,,\quad
    \adenom1^2\,\adenom3^2\,,\quad
    \adenom2^2\,\adenom3^2\,, 
\end{equation}
leading to the ansatz
\begin{equation}
\pe1\,\pe2\,\pe3^*\left(\frac{c_{12}}{\adenom1^2\,\adenom2^2} +\frac{c_{13}}{\adenom1^2\,\adenom3^2} +\frac{c_{23}}{\adenom2^2\,\adenom3^2}  \right)\,.
\end{equation}
We factored out the overall product of momenta and polarisations for simplicity.
We put the difference of the ansatz and the
expression~\eqref{eq:pole-seed} over a common denominator,
and then obtain equations by looking at all coefficients
of different numerator monomials in the $\adenom{i}$. Given $p \cdot \ell_3=p \cdot\left(\ell_1+\ell_2\right)$, this leads to
\begin{equation}
\begin{aligned}
c_{12}+c_{23}
-(\ell_1\cdot\ell_3)(\ell_2\cdot\ell_3)
+(\ell_1\cdot\ell_2)(\ell_1\cdot\ell_3)&=0,\\
c_{12}
-(\ell_1\cdot\ell_3)(\ell_2\cdot\ell_3)&=0,\\
c_{12}+c_{13}
-(\ell_1\cdot\ell_3)(\ell_2\cdot\ell_3)
+(\ell_1\cdot\ell_2)(\ell_2\cdot\ell_3)&=0.
\end{aligned}
\end{equation}
These equations can be solved easily, leading to
\begin{equation}
\begin{aligned}
c_{12}&=(\ell_1\cdot\ell_3)(\ell_2\cdot\ell_3),\\
c_{13}&=-(\ell_1\cdot\ell_2)(\ell_2\cdot\ell_3),\\
c_{23}&=-(\ell_1\cdot\ell_2)(\ell_1\cdot\ell_3).
\end{aligned}
\end{equation}
Other paths to similar equations (for example, using multivariate partial fractioning) may lead to more equations than unknowns, but the consistency of those equations is ultimately guaranteed by the existence of a worldline formulation.

We can already see the outlines of the vertex types which will emerge.
Each power of $\ell$ corresponds to a derivative.
The first power is the derivative inside $F_{\mu\nu}$ so that a lone power of $\ell$ will ultimately correspond to a single-$z$ vertex.
Additional powers will correspond to derivatives of $F$ (or of $z$) which come along with additional powers of $z$ as can be seen in the action~\eqref{eq:SchildSubtractedAction}.
These will correspond to multiple-$z$ vertices.

We now proceed to add terms to these coefficients in order to complete the ansatz into the amplitude, now organized in terms of worldline vertex structures.
As an example, take the $c_{12}$ coefficient.
We introduce Lorentz indices and pull apart the dot products so as to collect separately all objects associated with a given photon:
\begin{equation}
\pe1\,\pe2\,\pe3^* \, c_{12} = p_\alpha p_\beta p_\gamma \, (\polvector_1^\alpha \ell_1^\mu) \, (\polvector_2^\beta \ell_2^\nu ) \, (\polvector_3^{*\gamma} \ell_{3\mu} \ell_{3\nu}) \,.
\end{equation}
Now, to each term in parentheses, we add terms to reconstruct the Feynman vertices~\eqref{eq:SingleZVertex} and~\eqref{eq:twozvertex} with the $\omega_i$ replaced by $\pm\adenom{i}/m$. That is:
\begin{equation}
\begin{aligned}
\label{eq:replacements}
p_\alpha (\polvector_1^\alpha \ell_1^\mu)  &\to p_\alpha (\polvector_1^\alpha \ell_1^\mu -  \polvector_1^\mu \ell_1^\alpha) = p_\alpha \tilde{F}_1^{\mu\alpha} \,, \\
p_\beta (\polvector_2^\beta \ell_2^\nu)  &\to p_\beta (\polvector_2^\beta \ell_2^\nu -  \polvector_2^\nu \ell_2^\beta) = p_\beta \tilde{F}_2^{\nu\beta} \,, \\
p_\gamma (\polvector_3^{*\gamma} \ell_{3\mu} \ell_{3\nu}) &\to  p_\gamma (\polvector_3^{*\gamma} \ell_{3\mu} \ell_{3\nu} - \polvector_{3\mu}^* \ell_{1}^\gamma \ell_{3\nu} - \polvector_{3\nu}^* \ell_{3\mu} \ell_{2}^\gamma) \\
& \qquad = -\frac{1}{2} p^\gamma\left(i \partial _\mu \tilde{F}^3_{\nu \gamma}+i \partial_\nu \tilde{F}^3_{\mu \gamma} +\tilde{F}^3_{\nu \mu} \ell_{1\gamma} +\tilde{F}^3_{\mu \nu} \ell_{2\gamma}\right)\,.
\end{aligned}
\end{equation}
When we carry this out for all three coefficients, we find that we recover the full amplitude~\eqref{eq:fullAmp5}.
The factors that have the form $p \cdot \tilde{F}_i$, as in the first two lines of Eq.~\eqref{eq:replacements}, are candidates for subtraction when sewn into a loop amplitude.
The corresponding photon will require a subtraction when it is sewn onto a vertex consisting solely of the momentum of the other particle.

What about self-energy contributions?
In the worldline formalism, self-energy diagrams such as Fig.~\ref{fig:self-energy} are divergent and require regulation and appropriate subtraction.
At the level of the classical amplitude, we can cut and sew the self-energy photon in the same manner as the photons above.
Such a photon will require subtraction when it is sewn onto a vertex consisting solely of the momentum of the \emph{same} particle.

\providecommand{\href}[2]{#2}\begingroup\raggedright\endgroup

\end{document}